\documentclass[english]{article}
\usepackage[T1]{fontenc}
\usepackage[latin1]{inputenc}
\usepackage{babel}
\usepackage{makeidx}
\makeindex

\makeatletter

\providecommand{\LyX}{L\kern-.1667em\lower.25em\hbox{Y}\kern-.125emX\@}

 \usepackage{verbatim}
 \newcommand{\lyxaddress}[1]{
   \par {\raggedright #1 
   \vspace{1.4em}
   \noindent\par}
 }

\makeatother
\begin{document}

\title{Quantum Monte Carlo methods for the solution of the Schrödinger equation
for molecular systems}

\author{Alán Aspuru-Guzik and William A. Lester, Jr.}

\maketitle

\lyxaddress{Department of Chemistry, University of California, Berkeley, California
94720-1460}

\tableofcontents{}

\part{Preface}

The solution of the time independent Schr\"odinger equation for molecular
systems requires the use of modern computers, because analytic solutions
are not available.

This review deals with some of the methods known under the umbrella
term quantum Monte Carlo (QMC), specifically those that have been
most commonly used for electronic structure. Other applications of
QMC are widespread to rotational and vibrational states of molecules,
such as the work of \cite{DMBDCC00, YKPH+00, DCC01, AVKBW01}, condensed
matter physics \cite{DMC80,MFLM+01}, and nuclear physics \cite{RBW+00, SCP01}.

QMC methods have several advantages: 

\begin{itemize}
\item Computer time scales with system size roughly as \( N^{3} \), where
\( N \) is the number of particles of the system. Recent developments
have made possible the approach to linear scaling in certain cases.
\item Computer memory requirements are small and grow modestly with system
size.
\item QMC computer codes are significantly smaller and more easily adapted
to parallel computers than basis set molecular quantum mechanics codes.
\item Basis set truncation errors are absent in the QMC formalism.
\item Monte Carlo numerical efficiency can be arbitrarily increased. QMC
calculations have an accuracy dependence of \( \sqrt{T} \), where
\( T \) is the computer time. This enables one to choose an accuracy
range and readily estimate the computer time needed for performing
a calculation of an observable with an acceptable error bar. 
\end{itemize}
The purpose of the present work is to present a description of the
commonly used algorithms of QMC for electronic structure and to report
some recent developments in the field. 

The paper is organized as follows. In Sec. \ref{Sec:Introduction},
we provide a short introduction to the topic, as well as enumerate
some properties of wave functions that are useful for QMC applications.
In Sec. \ref{Sec:Algorithms} we describe commonly used QMC algorithms..
In Sec. \ref{Sec:SpecialTopics}, we briefly introduce some special
topics that remain fertile research areas. 

Other sources that complement and enrich the topics presented in this
chapter are our previous monograph, \cite{azul} and the reviews of
\cite{KES86, review:qmclester, review:qmcinchemistry, PHA97, DBPJR98, LM98, JBA99, ALJBA00, MFLM+01}.
There are also chapters on QMC contained in selected computational
physics texts \cite{SEK95, tobo, JMT99}. Selected applications of
the method are contained in Refs. \cite{CG96, ss00, app:silanos, HF94, jcg95, HJF97, CFCJU96, RNB, HJFCS97, JCG00, IVO+01}. 

QMC methods that are not covered in this review are the auxiliary
field QMC method \cite{DMC95, NRDMC97, RBMH98, RBDN00} and path integral
methods \cite{pimchelium, ASKES00}.

\index{atomic units}Atomic units are used throughout, the charge
of the electron, \( e \) and Planck's normalized constant, \( \hbar  \)
are set to unity. In this metric system, the unit distance is the
Bohr radius, \( a_{o} \).

\part{Introduction}

\label{Sec:Introduction}

The goal of the quantum Monte Carlo (QMC) method is to solve the Schr\"odinger
equation\index{Schr\"odinger equation}, which in the time independent
representation is given by,

\begin{equation}
\label{tdschroeq}
\hat{H}\Psi _{n}(\mathbf{R})=E_{n}\Psi _{n}(\mathbf{R})
\end{equation}

Here, \( \hat{H} \) is the Hamiltonian\index{Hamiltonian} operator
of the system in state \( n \), with wave function \( \Psi _{n}(\mathbf{R}) \)
and energy \( E_{n} \); \textbf{\( \mathbf{R} \)} is a vector that
denotes the \( 3N \) coordinates of the system of \( N \) particles
(electrons and nuclei), \( \mathbf{R}\equiv \{\mathbf{r}_{1},\ldots ,\mathbf{r}_{n}\} \).
For molecular systems, in the absence of electric or magnetic fields,
the Hamiltonian has the form \( \hat{H}\equiv \hat{T}+\hat{V} \)
, where \( \hat{T} \) is the kinetic energy\index{kinetic energy}
operator, \( \hat{T}\equiv -\frac{1}{2}\nabla ^{2}_{\mathbf{R}}\equiv -\frac{1}{2}\sum _{i}\nabla ^{2}_{i} \),
and \( \hat{V} \) is the potential energy\index{potential energy}
operator. For atomic and molecular systems \( \hat{V} \) is the Coloumb
potential\index{Coulomb interactions} between particles of charge
\( q_{i} \), \( \hat{V}\equiv \sum _{ij}\frac{q_{ij}}{\mathbf{r}_{ij}} \). 

The first suggestion of a Monte Carlo solution of the Schrödinger
equation dates back to Enrico Fermi, based on Metropolis, and Ulam
\cite{metulam}. He indicated that a solution to the stationary state
equation,

\begin{equation}
\label{simpleschroeq}
-\frac{1}{2}\nabla _{\mathbf{R}}^{2}\Psi (\mathbf{R})=E\Psi (\mathbf{R})-V(\mathbf{R})\Psi (\mathbf{R})
\end{equation}
could be obtained by introducing a wave function of the form, \( \Psi (\mathbf{R},\tau )=\Psi (\mathbf{R})e^{-E\tau } \).
This yields the equation,

\begin{equation}
\label{schrodiffsimple}
\frac{\partial \Psi (\mathbf{R},\tau )}{\partial \tau }=\frac{1}{2}\nabla ^{2}\Psi (\mathbf{R},\tau )-V(\mathbf{R})\Psi (\mathbf{R},\tau )
\end{equation}

Taking the limit \( \tau \to \infty  \), Eq. \ref{simpleschroeq}
is recovered. If the second term on the right hand side of Eq. \ref{schrodiffsimple}
is ignored, the equation is isomorphic with a diffusion equation\index{diffusion!equation},
which can be simulated by a random walk\index{random walk} \cite{AE26, RCKOF0},
where random walkers diffuse in a \( \mathbf{R} \)-dimensional space.
If the first term is ignored, the equation is a first-order kinetics
equation with a position dependent rate constant, \( V(\mathbf{R}) \),
which can also be interpreted as a stochastic survival probability\index{branching}.
A numerical simulation in which random walkers diffuse through \( \mathbf{R} \)-space,
reproduce in regions of of low potential, and die in regions of high
potential leads to a stationary distribution proportional to \( \Psi (\mathbf{R}) \),
from which expectation values can be obtained.

\section{Numerical solution of the Schrödinger equation}

\label{NumericalSolution}Most efforts to solve the Schrödinger equation
are wave function methods. These approaches rely exclusively on linear
combinations of Slater determinants, and include configuration interaction\index{configuration interaction (CI)}
(CI) and the multi-configuration self-consistent field \index{multi configuration self-consistent field (MCSCF)}(MCSCF).
There are perturbation approaches including the Möller-Plesset series
\index{MP(n) methods}(MP2, MP4), and coupled cluster (CC) \index{coupled cluster (CC)}
theory, which are presently popular computational procedures. Wave
function methods suffer from scaling deficiencies. An exact calculation
with a given basis set expansion requires \( N! \) computer operations,
where \( N \) is the the number of basis functions. Competitive methods
such as coupled cluster with singles and doubles, and triples perturbation
treatment, CCSD(T), scale as \( N^{7} \) %
\footnote{For a more detailed analysis of the scaling of wave function based
methods see, for example, \cite{MH96} and \cite{KRaJBA96}. For a
general overview of these methods, the reader is referred to Chapter
I of this book \cite{ECMD+02}.
}.

A term that we will use later is correlation energy\index{correlation energy}
(CE). It is defined as the difference between the exact non-relativistic
energy, and the energy of a mean field solution of the Schrödinger
equation, the Hartree-Fock\index{Hartree-Fock (HF)} method, in the
limit of an infinite basis set \cite{PL59, GSNHM94}, 

\begin{equation}
\label{correlation energy}
E_{corr}=E_{exact}-E_{HF},
\end{equation}
The CI, MCSCF, MP(N), and CC methods are all directed at generating
energies that approach \( E_{exact} \).

Other methods that have been developed include dimensional expansions
\cite{DWMD+96}, and the contracted Schrödinger equation method \cite{DAM98}.
For an overview of quantum chemistry methods, see Ref. \cite{PVRS+98}.

Since the pioneering work of the late forties to early sixties \cite{metulam, MDDaMK50, MHK62},
the MC and related methods have grown in interest. QMC methods have
an advantage with system size scaling, in the simplicity of algorithms
and in trial wave function forms that can be used.

\section{Properties of the exact wave function}

\textbf{\label{PropertiesExact}}The exact time independent wave function
solves the eigenvalue equation \ref{tdschroeq}. Some analytic properties
of this function are very helpful in the construction of trial functions
for QMC methods.

For the present discussion, we are interested in the discrete spectrum
of the \( \hat{H} \) operator. In most applications the total Schrödinger
equation \ref{tdschroeq} can be represented into an {}``electronic''
Schrödinger equation and a {}``nuclear'' Schrödinger equation based
on the large mass difference between electrons and nuclei; This is
the Born-Oppenheimer (BO) approximation\index{Born-Oppenheimer approximation (BO)}.
Such a representation need not be introduced in QMC but here is the
practical benefit of it that the nuclei can be held fixed for electronic
motion results in the simplest form of the electronic Schrödinger
equation.

The wave function also must satisfy the virial, hypervirial Hellman-Feynman\index{Hellman-Feynman theorem}
and generalized Hellman-Feynman theorems \cite{HH37, RPF39, MW78} %
\footnote{The Hellman-Feynman theorem is discussed in Sec. \ref{HellmanFeynmanSection}.
} . The local energy \cite{AREK60}, 

\begin{equation}
\label{elocal}
E_{L}(\mathbf{R})\equiv \frac{\hat{H}\Psi (\mathbf{R})}{\Psi (\mathbf{R})}
\end{equation}
 \index{local energy}is a constant for the exact wave function.

When charged particles meet, there is a singularity in the Coulomb
potential. This singularity must be compensated by a singularity in
the kinetic energy, which results in a discontinuity in the first
derivative, i. e., a \emph{cusp}\index{cusp conditions}, in the wave
function when two or more particles meet \cite{TK57, CRM+91}. For
one electron coalescing at a nucleus, if we focus in a one electron
function or orbital\index{orbital} \( \phi (\mathbf{r})=\chi (r)Y^{m}_{l}(\theta ,\phi ) \),
where \( \chi (r) \) is a radial function\index{radial function},
and \( Y^{m}_{l}(\theta ,\phi ) \) is a spherical harmonic\index{spherical harmonic}
with angular and magnetic quantum numbers \( l \) and \( m \), the
electron-nucleus cusp condition\index{cusp conditions!electron-nucleus}
is

\begin{equation}
\label{encusp}
\frac{1}{\eta (r)}\frac{d\eta (r)}{dr}\vert _{r=0}=-\frac{Z}{l+1},
\end{equation}

where \( \eta (r) \) is the radial wave function with the leading
\( r \) dependence factored out, \( \eta (r)\equiv \chi (r)/r^{m} \),,
and \( Z \) is the atomic number of the nucleus.

For electron-electron interactions, the cusp condition\index{cusp conditions!electron-electron}
takes the form,

\begin{equation}
\label{eecusp}
\frac{1}{\eta _{ij}(r)}\frac{d\eta _{ij}(r)}{dr}\vert _{r_{ij}=0}=\frac{1}{2(l+1)},
\end{equation}

where \( \eta _{ij}(r) \) is the \( r^{m} \) factored function for
the electron-electron radial distribution function.

Furthermore, \( \bar{\rho }(\mathbf{r}) \), the spherical average
of the electron density\index{electron!density}, \( \rho (\mathbf{r}) \) %
\footnote{If \( N \) is the number of electrons, then \( \rho (r) \) is defined
by

\[
\rho (r)=N\int \vert \Psi (\mathbf{R})\vert ^{2}d\mathbf{R}\]

}, must satisfy another cusp condition\index{cusp conditions!overall electron density},
namely,

\begin{equation}
\label{density_cusp}
\frac{\partial }{\partial r}\bar{\rho }(r)\vert _{r=0}=-2Z\bar{\rho }(r)
\end{equation}

at any nucleus. Another condition on \( \rho (\mathbf{r}) \) is that
asymptotically\index{wave function!asymptotic conditions}, it decays
exponentially\index{ionization potential}, 

\begin{equation}
\label{density_decay}
\rho (r\to \infty )\approx e^{-2\sqrt{2I_{0}}r}
\end{equation}

where \( I_{o} \) is the first ionization potential. This relation
can be derived from consideration of a single electron at large distance.
Details on these requirements can be found in refs. \cite{MMM75}
and \cite{ERD76}.

We discuss how to impose properties of the exact wave function on
QMC trial functions in Sec. \ref{properties_tf}.

\subsubsection{Approximate wave functions}

\label{approx_wf}James and Coolidge \cite{HHJ37} proposed three
accuracy tests of a trial wave function, \( \Psi _{T} \): the root
mean square error in \( \Psi _{T} \),

\begin{equation}
\label{definition_of_q}
\delta _{\Psi }=[\int (\Psi _{T}-\Psi _{o})^{2}d\mathbf{R}]^{\frac{1}{2}}
\end{equation}

the energy error,

\begin{equation}
\label{deltae}
\delta _{E}=E(\Psi _{T})-E_{o}
\end{equation}

and the root mean square local energy deviation,

\begin{equation}
\label{deltaelocal}
\delta _{E_{L}}=[\int \vert (\hat{H}-E_{o})\Psi \vert ^{2}d\mathbf{R}]^{\frac{1}{2}}
\end{equation}

where the local energy\index{local energy} is defined as in Eq. \ref{elocal}.

The calculation of \( \delta _{\Psi } \) by QMC requires sampling
the exact wave function, a procedure that will be described in Sec.
\ref{wave-noncommute}.

Several stochastic optimization schemes have been proposed for minimizing
expressions \ref{definition_of_q}-\ref{deltaelocal}. Most researchers
have focused on \ref{deltaelocal}, i.e., minimizing \( \delta _{E_{L}} \);
see, for example, McDowell \cite{McD81}. In Sec. \ref{wavefunction_optimization}
we turn to stochastic wave function optimization procedures.

\part{Algorithms}

\label{Sec:Algorithms}

In this section, we describe the computational procedures of QMC methods.
All of these methods use MC techniques widely used in other fields,
such as operations research, applied statistics, and classical statistical
mechanics simulations. Techniques such as importance sampling, correlated
sampling and MC optimization are similar in spirit to those described
in MC treatises \cite{WFB58,JMH64,JHH70,WWW76,McD81,kalosmontecarlo,mc:fishman,IMS94,IM99,ADNdF+01,JSL01}.
The reader is referred to the former for more details on the techniques
described in this Section.

We present the simple, yet powerful variational Monte Carlo\index{Monte Carlo!variational}
(VMC) method, in which the Metropolis MC\index{Monte Carlo!Metropolis} %
\footnote{This algorithm is also known as the M(RT)\( ^{2} \), due to the full
list of the authors that contributed to it's development, Metropolis,
Rosenbluth, Rosenbluth, Teller and Teller, see ref. \cite{mr2t2}. 
} method is used to sample a known trial function, \( \Psi _{T} \).
We follow with the projector Monte Carlo \index{Monte Carlo!projector}(PMC)
methods that sample the unknown ground state wave function.

\section{Variational Monte Carlo}

\subsection{Formalism}

Variational methods\index{Monte Carlo!variational} involve the calculation
of the expectation value of the Hamiltonian operator using a trial
wave function, \( \Psi _{T} \). This function is dependent on a set
of parameters, \( \Lambda  \), that are varied to minimize the expectation
value\index{expectation value}, i.e.,

\begin{equation}
\label{VMC_Rayleigh_Ritz}
\langle \hat{H}\rangle =\frac{\langle \Psi _{T}\vert \hat{H}\vert \Psi _{T}\rangle }{\langle \Psi _{T}\vert \Psi _{T}\rangle }\equiv E[\Lambda ]\geq E_{o}
\end{equation}

The expectation value \ref{VMC_Rayleigh_Ritz} can be sampled from
a probability distribution proportional to \( \Psi _{T}^{2} \), and
evaluated from the expression,

\begin{equation}
\label{vmc_elocal}
\frac{\int \mathbf{dR}\left[ \frac{\hat{H}\Psi _{T}(\mathbf{R})}{\Psi _{T}(\mathbf{R})}\right] \Psi ^{2}_{T}(\mathbf{R})}{\int \mathbf{dR}\Psi ^{2}_{T}(\mathbf{R})}\equiv \frac{\int \mathbf{dR}E_{L}\Psi ^{2}_{T}(\mathbf{R})}{\int \mathbf{dR}\Psi ^{2}_{T}(\mathbf{R})}\geq E_{o},
\end{equation}
where \( E_{L} \) is the local energy defined in Sec. \ref{approx_wf}.
The procedure involves sampling random points in \( \mathbf{R} \)-space
from,

\begin{equation}
\label{p_r_vmc}
\mathcal{P}(\mathbf{R})\equiv \frac{\Psi _{T}^{2}(\mathbf{R})}{\int \mathbf{dR}\Psi ^{2}_{T}(\mathbf{R})}
\end{equation}

The advantage of using \ref{p_r_vmc} as the probability density function\index{probability density function (pdf)}
is that one need not perform the averaging of the numerator and denominator
of Eq. \ref{vmc_elocal}. The calculation of the ratio of two integrals
with the MC method is biased by definition: the average of a quotient
is not equal to the quotient of the averages, so this choice of \( P(\mathbf{R}) \)
avoids this problem.

In general, sampling is done using the Metropolis method \index{Monte Carlo!Metropolis}\cite{mr2t2},
that is well described in Chapter 3 of \cite{kalosmontecarlo}, and
briefly summarized here in Sec. \ref{generalized_metropolis_section}.

Expectation values\index{expectation value} can be obtained using
the VMC method from the following general expressions \cite{DBPJR99},

\begin{equation}
\label{coordinate_operator}
\langle \hat{O}\rangle \equiv \frac{\int \mathbf{dR}\Psi _{T}(\mathbf{R})^{2}\hat{O}(\mathbf{R})}{\int \mathbf{dR}\Psi _{T}(\mathbf{R})^{2}}\cong \frac{1}{N}\sum ^{N}_{i=1}\hat{O}(\mathbf{R}_{i}),
\end{equation}

\begin{equation}
\label{differential operator}
\langle \hat{O}_{d}\rangle \equiv \frac{\int \mathbf{dR}\left[ \frac{\hat{O}_{d}\Psi _{T}(\mathbf{R})}{\Psi _{T}(\mathbf{R})}\right] \Psi _{T}(\mathbf{R})^{2}}{\int \mathbf{dR}\Psi _{T}(\mathbf{R})^{2}}\cong \frac{1}{N}\sum ^{N}_{i=1}\frac{\hat{O}_{d}\Psi _{T}(\mathbf{R}_{i})}{\Psi _{T}(\mathbf{R}_{i})}.
\end{equation}
Equation \ref{coordinate_operator} is for a coordinate operator,
\( \hat{O} \), and \ref{differential operator} is preferred for
a differential operator, \( \hat{O}_{d} \).

\subsubsection{The generalized Metropolis algorithm}

\label{generalized_metropolis_section}\index{Monte Carlo!Metropolis}

The main idea of the Metropolis algorithm is to sample the electronic
density, given hereby, \( \Psi ^{2}_{T}(\mathbf{R}) \) using fictitious
kinetics\index{dynamics!fictitious} that in the limit of large simulation
time yields the density at equilibrium. A coordinate move is proposed,
\( \mathbf{R}\to \mathbf{R}' \), which has the probability of being
accepted\index{acceptance probability} given by,

\begin{equation}
\label{transition}
P(\mathbf{R}\to \mathbf{R}')=\min \left( 1,\frac{T(\mathbf{R}'\to \mathbf{R})\Psi _{T}^{2}(\mathbf{R}')}{T(\mathbf{R}\to \mathbf{R}')\Psi ^{2}_{T}(\mathbf{R})}\right) ,
\end{equation}
where \( T(\mathbf{R}\to \mathbf{R}') \) denotes the transition probability\index{transition probability}
for a coordinate move from \( \mathbf{R} \) to \textbf{\( \mathbf{R}' \)}.
In the original Metropolis procedure, \( T \) was taken to be a uniform
random distribution over a coordinate interval \( \Delta \mathbf{R} \).
Condition \ref{transition} is necessary to satisfy the detailed balance
\emph{}condition\emph{\index{detailed balance},}

\begin{equation}
\label{detailed_balance}
T(\mathbf{R}'\to \mathbf{R})\Psi ^{2}_{T}(\mathbf{R}')=T(\mathbf{R}\to \mathbf{R}')\Psi ^{2}_{T}(\mathbf{R})
\end{equation}

which is necessary for \( \Psi ^{2}_{T}(\mathbf{R}) \) to be the
equilibrium distribution of the sampling process.

Several improvements to the Metropolis method have been pursued both
in classical and in QMC simulations. These improvements involve new
transition probability functions and other sampling procedures. See,
for example, \cite{MHK74, DCGVC77, MRBJB79, CPMR79, DMC80, qmc:ha, DBPJR99, MD00b}. 

A common approach for improving \( T(\mathbf{R}\to \mathbf{R}') \)
in VMC, is to use the quantum force\index{quantum force}, 

\begin{equation}
\label{quantum_force}
\mathbf{F}_{\mathbf{q}}\equiv \nabla \ln \vert \Psi _{T}(\mathbf{R})^{2}\vert 
\end{equation}

as a component of the transition probability. The quantum force can
be incorporated by expanding \( f(\mathbf{R},\tau )=\vert \Psi _{T}(\mathbf{R})^{2}\vert =e^{-\ln \vert \Psi ^{2}_{T}(\mathbf{R})\vert } \),
in a Taylor series in \( \ln \vert \Psi ^{2}_{T}(\mathbf{R})\vert  \),
and truncating at first order, 

\begin{equation}
\label{tprob_force_vmc}
T(\mathbf{R}\to \mathbf{R}')\approx \frac{1}{N}e^{\lambda \mathbf{F}_{\mathbf{q}}(\mathbf{R})\cdot (\mathbf{R}'-\mathbf{R})},
\end{equation}

where \( N \) is a normalization factor, and \( \lambda  \) is a
parameter fixed for the simulation or optimized in some fashion, for
example, see \cite{MLS98}. A usual improvement is to introduce a
cutoff in \( \Delta \mathbf{R}=(\mathbf{R}'-\mathbf{R}) \), so that
if the proposed displacement is larger than a predetermined measure,
the move is rejected.

A good transition probability should also contain random displacements,
so that all of phase space can be sampled. The combination of the
desired drift arising from the quantum force of Eq. \ref{tprob_force_vmc}
with a Gaussian random move, gives rise to Langevin fictitious dynamics\index{dynamics!fictitious!Langevin},
namely,

\begin{equation}
\label{langevin_dynamics}
\mathbf{R}^{'}\to \mathbf{R}+\frac{1}{2}\mathbf{F}_{\mathbf{q}}(\mathbf{R})+\mathcal{G}_{\delta \tau },
\end{equation}

where \( \mathcal{G}_{\delta \tau } \) is a number sampled from a
Gaussian distribution\index{Gaussian distribution} with standard
deviation \( \delta \tau  \). The propagator\index{propagator} or
transition probability for Eq. \ref{langevin_dynamics} is 

\begin{equation}
\label{t_prob_fokker_planck}
T_{L}(\mathbf{R}\to \mathbf{R}')=\frac{1}{\sqrt{4\pi D\delta \tau }^{3N}}e^{-(\mathbf{R}'-\mathbf{R}-\frac{1}{2}\mathbf{F}_{\mathbf{q}}(\mathbf{R})\delta \tau )^{2}/2\delta \tau }
\end{equation}

which is a drifting Gaussian, spreading in \( \delta \tau  \). Using
Eq. \ref{langevin_dynamics} is equivalent to finding the solution
of the Fokker-Planck\index{Fokker-Planck equation} equation \cite{RCKOF0},

\begin{equation}
\label{fokker_plank_eq}
\frac{\partial f(\mathbf{R},\tau )}{\partial \tau }=\frac{1}{2}\nabla \cdot (\nabla -\mathbf{F}_{\mathbf{q}})f(\mathbf{R},\tau ),
\end{equation}

Equation \ref{t_prob_fokker_planck} has proved to be a simple and
effective choice for a VMC transition probability. More refined choices
can be made, usually with the goal of increasing acceptance probabilities
in regions of rapid change in \( \vert \Psi _{T}(\mathbf{R})^{2}\vert  \),
such as close to nuclei. For a more detailed discussion of this formalism,
the reader is directed to Chapter 2 of \cite{azul}. More elaborate
transition rules\index{transition probability} can be found in \cite{CJU93, ZSMMS94, MLS98, DBPJR99b}.

\subsubsection{Statistics}

\label{Statistics}\index{blocking statistics}Usually, VMC calculations
are performed using an ensemble of \( N_{\mathcal{W}} \) random walkers\index{random walker!ensemble},
\( \mathcal{W}\equiv \{\mathbf{R}_{1},\mathbf{R}_{2},\ldots ,\mathbf{R}_{\mathbf{n}}\} \)
that are propagated following \( T(\mathbf{R}\to \mathbf{R}') \)
using the probability \( P(\mathbf{R}\to \mathbf{R}') \) to accept
or reject proposed moves for ensemble members. Statistical averaging
has to take into account auto-correlation between moves that arises
if the mean square displacement\index{displacement!mean square} for
the ensemble, \( \Delta (\mathbf{R}\to \mathbf{R}')^{2}/N_{\mathcal{W}} \),
is sufficiently large. In such cases, observables measured at the
points \( \mathbf{R}' \) will be statistically correlated with those
evaluated at \( \mathbf{R} \). The variance\index{variance} for
an observable, \( \hat{O} \), measured over \( N_{s} \) MC steps\index{time step}
of a random walk is,

\begin{equation}
\label{variance_obs}
\sigma _{\hat{O}}\equiv \frac{1}{N_{s}N_{\mathcal{W}}}(O_{i}-\langle O\rangle ),
\end{equation}

where \( \langle O\rangle  \) is the average of the observations,
\( O_{i} \) over the sample. A simple approach to remove auto-correlation\index{auto-correlation}
between samples is to define a number of blocks, \( N_{b} \), where
each block is an average of of \( N_{s} \) steps, with variance,

\begin{equation}
\label{variance_block}
\sigma _{B}\equiv \frac{1}{N_{B}N_{\mathcal{W}}}(O_{b}-\langle O\rangle ),
\end{equation}

where \( O_{b} \) is the average number of observations \( N_{t} \)
in block \textbf{\( b \).} If \( N_{t} \) is sufficiently large,
\( \sigma _{B} \) is a good estimator of the variance of the observable
over the random walk. The auto-correlation time\index{auto-correlation!time}
is a good measure of computational efficiency, and is given by

\begin{equation}
\label{t_corr}
T_{corr}=\lim _{N_{s}\to \infty }N_{s}\left( \frac{\sigma ^{2}_{B}}{\sigma ^{2}_{\hat{O}}}\right) 
\end{equation}

The efficiency of a method depends on time step\index{time step}
\cite{SMR88}. Serial correlation between sample points should vanish
for an accurate estimator of the variance. For an observable \( \langle O\rangle  \)
, the serial correlation coefficient is defined as,

\textbf{\begin{equation}
\label{corrcoef}
\xi _{k}\equiv \frac{1}{(\langle O^{2}\rangle -\langle O\rangle ^{2})(N-k)}\sum ^{N-k}_{i=1}(O_{i}-\langle O\rangle )(O_{i+k}-\langle O\rangle ),
\end{equation}
}where \( k \) \textbf{}is the number of MC steps between the points
\( O_{i} \) and \( O_{i+k} \). \textbf{}The function \ref{corrcoef}
decays exponentially with \( k \). The correlation length, \( L, \)
is defined as the number of steps necessary for \( \xi _{k} \) to
decay essentially to zero. For an accurate variance estimator, blocks
should be at least \( L \) steps long. 

The efficiency\index{simulation efficiency} of a simulation is inversely
proportional to \( \xi _{k} \). The \( \xi _{k} \) dependence on
time step is usually strong \cite{azul}; the larger the time step,
the fewer steps/block \( L \) necessary, and the more points available
for calculating the global average \( \langle O\rangle  \). A rule
of thumb is to use an \( N_{t} \) \( \approx 10 \) times larger
than the auto-correlation time to insure statistical independence
of block averages, and therefore a reliable variance estimate.

The VMC method shares some of the strengths and weaknesses of traditional
variational methods: the energy is an upper bound to the true ground
state energy. If reasonable trial functions are used, often reliable
estimates of properties can be obtained. For quantum MC applications,
VMC can be used to obtain valuable results. In chemical applications,
VMC is typically used to analyze and generate trial wave functions
for PMC.

\subsection{Trial wave functions}

In contrast to wave function methods, where the wave function is constructed
from linear combinations of determinants of orbitals, QMC methods
can use arbitrary functional forms for the wave function subject to
the requirements in Sec. \ref{PropertiesExact}. Because QMC trial
wave functions are not restricted to expansions in one-electron functions
(orbitals), more compact representations are routinely used. In this
section, we review the forms most commonly used for QMC calculations.

\index{trial  function}\label{kinds_of_tf} Fermion wave functions
must be antisymmetric with respect to the exchange of an arbitrary
pair of particle coordinates. If they are constructed as the product
of \( N \) functions of the coordinates, \( \phi (r_{1},r_{2},\ldots r_{N}) \),
the most general wave function can be constructed enforcing explicit
permutation\index{permutation},

\begin{equation}
\label{antisymmetry_requirements}
\Psi (\mathbf{R},\Sigma )=\frac{1}{\sqrt{(N\cdot M)!}}\sum _{n,m}(-1)^{n}\hat{S}_{m}\hat{P}_{n}\phi (r_{1},\sigma _{1}r_{2},\sigma _{2},\ldots ,r_{N},\sigma _{N}),
\end{equation}

where \( \hat{P}_{n} \) is the \( n^{th} \) coordinate permutation
operator, \( \hat{P}_{n}\phi (r_{1},r_{2},\ldots r_{i},r_{j},\ldots r_{N})=\phi (r_{1},r_{2},\ldots r_{j},r_{i},\ldots r_{N}) \),
and \( \hat{S}_{m}\phi (\sigma _{1},\sigma _{2},\sigma _{i},\sigma _{j}\ldots ,\sigma _{N})=\phi (\sigma _{1},\sigma _{2},\sigma _{j},\sigma _{i}\ldots ,\sigma _{N}) \)
is the \( m^{th} \) spin\index{spin} coordinate permutation operator.

If the functions \( \phi _{i} \) depend only on single-particle coordinates,
their antisymmetrized product can be expressed as a Slater determinant,

\begin{equation}
\label{determinante}
D(\mathbf{R},\Sigma )=\frac{1}{\sqrt{N!}}\det \vert \phi _{1},\ldots ,\phi _{i}(r_{j},\sigma _{j}),\ldots ,\phi _{n}\vert 
\end{equation}

Trial wave functions constructed from orbitals scale computationally
as \( N^{3} \), where \( N \) is the system size\index{determinant},
compared to \( N! \) for the fully antisymmetrized form %
\footnote{The evaluation of a determinant of size \( N \) requires \( N^{2} \)
computer operations. If the one-electron functions scale with system
size as well, the scaling becomes \( N^{3} \)\label{foot_comment}.
In contrast, a fully antisymmetrized form requires the explict evaluation
of the \( N! \) permuations, making the evaluation of this kind of
wave functions in QMC prohibitive for systems of large \( N \).
}. The number of evaluations can be reduced by determining which permutations
contribute to a particular spin state.

For QMC evaluation of properties that do not depend on spin coordinates\index{spin!coordinates},
\( \Sigma  \), for a given spin state, the \( M! \) configurations
that arise from relabeling electrons, need not be evaluated. The reason
is that the Hamiltonian of Eq. \ref{tdschroeq}, contains no magnetic
or spin operators and spin degrees of freedom remain unchanged. In
this case, and for the remainder of this paper, \( \sigma _{\uparrow } \)
electrons do not permute with \( \sigma _{\downarrow } \)electrons,
so that the full Slater determinant(s) can be factored into a product
of spin-up, \( D^{\uparrow } \) and spin-down, \( D^{\downarrow } \)
determinants. The number of allowed permutations is reduced from \( (N_{\uparrow }+N_{\downarrow })! \)
to \( N_{\uparrow }!N_{\downarrow }! \) \cite{dmcpuro:caffarel, SHZS90}.

\label{properties_tf}

The use of various wave function forms in QMC has been explored by
\cite{SAARLC97} , as well as \cite{dario:ce}. Fully antisymmetric
descriptions of the wave function are more flexible and require fewer
parameters than determinants, but their evaluation is inefficient
due to the \( N! \) scaling.

A good compromise is to use a product wave function of a determinant
or linear combination of determinants, e.g., HF\index{Hartree-Fock},
MCSCF\index{MCSCF}, CASSCF\index{CASSCF}, CI\index{configuration interaction (CI)},
multiplied by a correlation function that is symmetric with respect
to particle exchange\index{pair-product wave function},

\begin{equation}
\label{pair_product}
\Psi _{T}=\mathcal{DF},
\end{equation}

Here \( \mathcal{D} \) denotes the antisymmetric wave function factor
and \( \mathcal{F} \) is the symmetric factor. We now describe some
of the forms used for \( \mathcal{D} \) and then we describe forms
for \( \mathcal{F} \). Such products are also known as the correlated
molecular orbital\index{wave function!correlated molecular orbital}
(CMO) wave function.

In the CMO wave functions, the antisymmetric part of the wave function
is constructed as a determinant of independent particle functions,
\( \phi _{i} \) (see Eq. \ref{determinante}). The \( \phi _{i} \)
are usually formed as a linear combination of basis functions centered
on atomic centers, \( \phi _{i}=\sum _{j}c_{j}\chi _{j} \). The most
commonly used basis functions in traditional \emph{ab initio} quantum
chemistry are Gaussian functions\index{basis function!Gaussian},
which owe their popularity to ease of integration of molecular integrals.
Gaussian basis functions take the form,

\begin{equation}
\label{gaussian_mo}
\chi _{G}\equiv x^{a}y^{b}z^{c}e^{-\xi \mathbf{r}^{2}}
\end{equation}

For QMC applications, it is better to use the Slater-type\index{basis function!Slater}
basis functions,

\begin{equation}
\label{slater_mo}
\chi _{S}\equiv x^{a}y^{b}z^{c}e^{-\xi \mathbf{r}}
\end{equation}

because they rigorously satisfy the electron-nuclear cusp\index{cusp conditions!electron-nucleus}
condition (see Eq. \ref{encusp}), and the asymptotic property\index{wave function!asymptotic conditions}
of Eq. \ref{density_decay}. Nevertheless, in most studies, Gaussian
basis functions have been used, and corrections for enforcing the
cusp conditions can be made to improve local behavior close to a nucleus.
For example, in one approach \cite{SMAL01}, the region close to a
nucleus is described by a Slater-type function, and a polynomial fit
is used to connect the Gaussian region to the exponential. This procedure
strongly reduces fluctuations of the kinetic energy of these functions,
a desirable property for guided VMC and Green's function methods.

The symmetric part of the wave function is usually built as a product
of terms explicitly dependent on inter-particle distance, \textbf{\( \mathbf{r}_{ij}=\vert \mathbf{r}_{i}-\mathbf{r}_{j}\vert  \).}
These functions are usually constructed to reproduce the form of the
wave function at electron-electron and electron-nucleus cusps. A now
familiar form is that proposed by \cite{AB40, RBD49, jastrow}, and
known as the Jastrow \emph{ansatz.}\index{correlation function!Jastrow},

\begin{equation}
\label{jastrow}
\mathcal{F}\equiv e^{U(\mathbf{r}_{ij})}\equiv e^{\prod _{i<j}g_{ij}},
\end{equation}

where the correlation function \( g_{ij} \) is,

\begin{equation}
\label{jastrow_gij}
g_{ij}\equiv \frac{a_{ij}\mathbf{r}_{ij}}{1+b_{ij}\mathbf{r}_{ij}}
\end{equation}

with constants specified to satisfy the cusp conditions\index{cusp conditions},

\begin{equation}
\label{aij}
a_{ij}\equiv \left\{ \begin{array}{ll}
\frac{1}{4}, & \mathrm{if}\, ij\, \mathrm{are}\, \mathrm{like}\, \mathrm{spins}\\
\frac{1}{2}, & \mathrm{if}\, ij\, \mathrm{are}\, \mathrm{unlike}\, \mathrm{spins}\\
1, & \mathrm{if}\, ij\, \mathrm{are}\, \mathrm{electron}/\mathrm{nucleus}\, \mathrm{pairs}
\end{array}\right. 
\end{equation}

Electron correlation for parallel spins is taken into account by the
Slater determinant. 

This simple Slater-Jastrow\index{wave function!Slater-Jastrow} \emph{ansatz}
has a number of desirable properties. First, as stated above, scaling
with system size for the evaluation of the trial function is \( N^{3} \)\index{trial function!scaling},
where \( N \) is the number of particles in the system, the correct
cusp conditions are satisfied at two-body coalescence points and the
correlation function \( g_{ij} \) approaches a constant at large
distances, which is the correct behavior as \( \mathbf{r}_{ij}\to \infty  \).

In general, the inclusion of 3-body and 4-body correlation terms has
been shown to improve wave function quality. The work of \cite{CHCJU97}
shows that if the determinant parameters \( \lambda _{D} \) are optimized
along with the correlation function parameters, \( \lambda _{C} \),
one finds that the nodal structure of the wave function does not improve
noticeably in going from 3- to 4-body correlation terms, which suggests
that increasing the number of determinants, \( N_{D} \) is more important
than adding fourth- and higher-order correlation terms.

The use of Feynman-Cohen backflow\index{backflow effects} correlations
\cite{RPF56}, which has been suggested \cite{KES90} for the inclusion
of three body correlations in \( U \), has been used in trial functions
for homogeneous systems such as the electron gas \cite{YKDMC93, YKDMC98}
and liquid helium \cite{KESKAL80, JCJB00}. Feynman \cite{RPF56}
suggested replacing the orbitals by functions that include hydrodynamic
\emph{backflow} effects. His idea was based on the conservation of
particle current and the variational principle. The procedure involves
replacing mean field orbitals by \emph{backflow-}corrected orbitals
of the form,

\begin{equation}
\label{backflow_orbitals}
\phi _{n}(\mathbf{r}_{i})\to \phi _{n}(\mathbf{r}_{i}+\sum _{j\neq i}\mathbf{r}_{ij}\nu (\mathbf{r}_{ij})),
\end{equation}

where \( \nu (\mathbf{r}_{ij}) \) is the backflow function. Others
\cite{VRP73} proposed that \( \nu (r_{ij}) \) should consist of
the difference between the \( l=0 \) and \( l=1 \) states of an
effective two-particle Schrödinger equation. Furthermore, they proposed
\cite{VRP86} the inclusion of a \( 1/r^{3} \) tail, as originally
suggested by Feynman and Cohen,

\begin{equation}
\label{correlation}
\nu (\mathbf{r})=\lambda _{\nu }e^{-\left[ \frac{\mathbf{r}_{i}-\mathbf{r}_{j}}{\omega _{\nu }}\right] ^{2}}+\frac{\lambda _{\nu '}}{\mathbf{r}^{3}},
\end{equation}

where, \( \lambda _{\nu } \), \( \lambda _{\nu '} \), and \( \omega _{\nu } \)
are variational parameters. As recently noted \cite{YKDMC93}, the
incorporation of the full backflow trial function into wave functions
involves a power of \( N \) increase in computational expense, but
yields a better DMC energy for the electron gas\index{electron!gas}%
\footnote{As discussed in Sec. \ref{Section:fixednode}, an improved fixed-node
energy is a consequence of better nodes of the trial wave function,
a critically important characteristic for importance sampling functions
in QMC methods.
}\index{electron!gas}.

Recently, one has seen the practice of taking orbitals from a mean
field calculation and the inclusion of averaged backflow terms in
the correlation function \( \mathcal{F} \). The advantage of this
approach is that orbitals are unperturbed and readily obtainable from
mean field computer codes.

The correlation function form used by \cite{KES90} is a selection
of certain terms of the general form originally proposed in connection
with the transcorrelated method\cite{SFB69}\index{correlation function!Boys-Handy},

\begin{equation}
\label{Boys-Handy-sum}
\mathcal{F}=e^{\sum _{I,i<j}U_{Iij}},
\end{equation}

where

\begin{equation}
\label{Boys-Handy}
U_{Iij}=\sum _{k}^{N(I)}\Delta (m_{kI}n_{kI})c_{kI}(g^{m_{kI}}_{iI}g^{n_{kI}}_{jI}+g^{m_{kI}}_{jI}g^{n_{kI}}_{iI})g^{o_{kI}}_{ij}.
\end{equation}

The sum in \ref{Boys-Handy-sum} goes over \( I \) nuclei, \( ij \)
electron pairs, and the sum in Eq. \ref{Boys-Handy} is over the \( N(I) \)
terms of the correlation function for each nucleus. The parameters
\( m,n \) and \( o \) are integers. The function \( \Delta (m,n) \)
takes the value \( 1 \) when \( m\neq n \), and \( \frac{1}{2} \)
otherwise. The functions \( g_{ij} \) are specified by Eq. \ref{jastrow_gij}.

This correlation function \ref{Boys-Handy-sum}, \ref{Boys-Handy}
can be shown to have contributions to averaged backflow effects from
the presence of electron-electron-nucleus correlations that correspond
to values of \( m,n \) and \( o \) in Eq. \ref{Boys-Handy} of 2,2,0
and 2,0,2. These contributions recover \( \approx 25\% \) or more
of the total correlation energy\index{correlation energy} of atomic
and molecular systems above that from the simple Jastrow term \cite{KES90}.

\subsection{The variational Monte Carlo algorithm}

\index{Monte Carlo!variational!algorithm}\label{vmc_algorithm}

The VMC algorithm is an application of the generalized Metropolis
MC method. As in most applications of the method, one needs to insure
that the ensemble has achieved equilibrium in the simulation sense.
Equilibrium is reached when the ensemble \( \mathcal{W} \) is distributed
according to \( \mathcal{P}(\mathbf{R}) \) This is usually achieved
by performing a Metropolis random walk and monitoring the trace of
the observables of interest. When the trace fluctuates around a mean,
it is generally safe to start averaging in order to obtain desired
properties.

An implementation of the VMC algorithm follows:

\begin{enumerate}
\item Equilibration stage\index{equilibration}

\begin{enumerate}
\item Generate an initial set of random walker positions, \( \mathcal{W}_{o} \);
it can be read in from a previous random walk, or generated at random.
\item Perform a loop over \( N_{s} \) steps,

\begin{enumerate}
\item For each \( r_{i} \) of the \( N_{p} \) number of particles,

\begin{enumerate}
\item Propose a move from \( \Psi (\mathbf{R})\equiv \Psi (\mathbf{r}_{1},\mathbf{r}_{2},\ldots ,\mathbf{r}_{i},\ldots ,\mathbf{r}_{N_{p}}) \)
to \( \Psi (\mathbf{R}')\equiv \Psi (\mathbf{r}_{1},\mathbf{r}_{2},\ldots ,\mathbf{r}'_{i},\ldots ,\mathbf{r}_{N_{p}}) \).
Move from \textbf{\( \mathbf{r} \)} to \textbf{\( \mathbf{r}' \)}
according to\\
\begin{equation}
\label{vmc_guided_t}
\mathbf{r}'\leftarrow \mathbf{r}+\mathcal{G}_{\delta \tau }+\frac{1}{2}\mathbf{F}_{\mathbf{q}}\delta \tau ,
\end{equation}
where \( \mathcal{G}_{\delta \tau } \) is a Gaussian random number
with standard deviation \( \delta \tau  \),which is a proposed step
size, and \( \mathbf{F}_{\mathbf{q}} \) is the quantum force (see
Eq. \ref{tprob_force_vmc}). This is the Langevin dynamics of Eq.
\ref{langevin_dynamics}\index{dynamics!fictitious!Langevin}.
\item Compute the Metropolis acceptance/rejection\index{acceptance probability}
probability,\\
\begin{equation}
\label{transition2}
P(\mathbf{R}\to \mathbf{R}')=\min \left( 1,\frac{T_{L}(\mathbf{R}'\to \mathbf{R})\Psi _{T}^{2}(\mathbf{R}')}{T_{L}(\mathbf{R}\to \mathbf{R}')\Psi ^{2}_{T}(\mathbf{R})}\right) ,
\end{equation}
where \( T_{L} \) is given by Eq. \ref{t_prob_fokker_planck}.
\item Compare \( P(\mathbf{R}\to \mathbf{R}') \) with an uniform random
number between 0 and 1, \( \mathcal{U}_{[0,1]} \). If \( P>\mathcal{U}_{[0,1]} \),
accept the move, otherwise, reject it.
\item Calculate the contribution to the averages \( \frac{\hat{O}_{d}\Psi _{T}(\mathbf{R}')}{\Psi _{T}(\mathbf{R}')} \),
and perform blocking statistics as described in Sec. \ref{Statistics}.
\end{enumerate}
\item Continue the loop until the desired accuracy is achieved.
\end{enumerate}
\end{enumerate}
\end{enumerate}

\section{Wave function optimization}

\label{wavefunction_optimization} 

Trial wave functions \( \Psi _{T}(\mathbf{R},\Lambda ) \) for QMC
are dependent on variational parameters,\index{trial function!parameter optimization}
\( \Lambda =\{\lambda _{1},\ldots ,\lambda _{n}\} \). Optimization
of \( \Lambda  \) is a key element for obtaining accurate trial functions.
Importance sampling using an optimized trial function increases the
efficiency of DMC simulations. There is a direct relationship between
trial-function accuracy and the computer time required to calculate
accurate expectation values. Some of the parameters \( \lambda _{i} \)
may be fixed by imposing appropriate wave function properties, such
as cusp conditions (See Sec. \ref{PropertiesExact}).

It is useful to divide \( \Lambda  \) into groups distinguished by
whether the optimization changes the nodes of the wave function. The
Slater determinant parameters, \( \lambda _{\mathcal{D}^{\uparrow \downarrow }} \)
and the Slater determinant weights, \( \lambda _{k_{i}} \)change
wave function nodal structure\index{fermion nodes} \cite{umrigar:opttf, KES90, ZSRNB92, HJFCS97, opt:pdf, RNB}.
The correlation function\index{correlation function!parameters} parameters,
\( \lambda _{\mathcal{F}} \) do not change the nodal structure of
the overall wave function, and therefore the DMC energy. For some
systems, the optimization of \( \lambda _{\mathcal{F}} \) is sufficient
for building reliable trial functions for PMC methods, because \( \mathcal{F} \)
is designed in part to satisfy cusp conditions \cite{TK57, CRM+91, CFCJU96}\index{cusp conditions}.

There have been several optimization methods proposed previously.
Some involve the use of analytical derivatives \cite{SHZS90, HBSMR92, ALJBA96, HHZC96, HHQX+99, XLHZ00},
and others focus on the use of a fixed sample for variance minimization
Conroy \cite{HC64}, and more recently others \cite{umrigar:opttf, ZSSH+90, MPN97}.
Yet another direction is the use of histogram analysis for optimizing
the energy, variance, and molecular geometry for small systems \cite{MSJRD99}.
In the present study, we concentrate on fixed sample optimization
to eliminate stochastic uncertainty during the random walk \cite{ZSSH+90}.

Other authors optimize the trial wave function using information obtained
from a DMC random walk \cite{CFSF00}. This approach shows promise,
because usually the orbitals obtained from a mean field theory, such
as HF or LDA, are frozen and used in the DMC calculation without re-optimization
specifically for correlation effects within the DMC framework.

The common variance functional (VF) \index{variance functional (VF)}\cite{umrigar:opttf}
is given by

\begin{equation}
\label{vf}
VF=\frac{\sum ^{N}_{i=1}\left[ \frac{\hat{H}\Psi (\mathbf{R}_{i},\Lambda )}{\Psi (\mathbf{R}_{i},\Lambda )}-E_{T}\right] ^{2}w_{i}}{\sum ^{N}_{i=1}w_{i}},
\end{equation}
where \( E_{T} \) is a trial energy, \( w_{i} \) is a weighting
factor defined by

\begin{equation}
\label{WEIGTH}
w_{i}(\Lambda )=\frac{\Psi _{i}^{2}(\mathbf{R}_{i},\Lambda )}{\Psi _{i}^{2}(\mathbf{R}_{i},\Lambda _{0})},
\end{equation}
and \( \Lambda _{0} \) is an initial set of parameters. The sum in
Eq. \ref{vf} is over fixed sample configurations.

\subsubsection{Trial wave function quality\index{wave function!quality}}

The overlap of \( \Psi _{T} \) with the ground state wave function,
\( \langle \Psi _{T}\vert \Psi _{0}\rangle  \), by DMC methods,\cite{MHMS00b}
is an very efficient way of assessing wave function quality. There
is also a trend that correlates the variational energy of the wave
function with associated variance in a linear relationship \cite{YKDMC93,YKDMC98}.
This correlation is expected because both properties, \( \delta _{E} \),
and \( \delta _{E_{L}} \), approach limits -- \( E_{o} \) and zero
respectively -- as wave function quality improves. Observing this
correlations is a good method of validating the optimization method,
as well as assessing wave function quality.

\section{Projector methods}

\index{Monte Carlo!projector}\label{GreenFChapter}

QMC methods such as DMC and GFMC are usefully called projector Monte
Carlo (PMC) methods %
\footnote{The introductory section of this Chapter, follows the work of \cite{JHH84, SAC90, NJC95}.
}. The general idea is to project out a state of the Hamiltonian by
iteration of a projection operator\index{projection operator}, \( \hat{P} \).
For simplicity, we assume that the desired state is the ground state,
\( \Psi _{o} \), but projectors can be constructed for any state,

\begin{equation}
\label{gstate}
\lim _{i\to \infty }\hat{P}^{i}\vert \Psi _{T}\rangle \approx \vert \Psi _{0}\rangle .
\end{equation}

After sufficient iterations \( i \), the contribution of all excited
states \( \vert \Psi _{i}\rangle  \), will be filtered out, and only
the ground state is recovered.

If \( \vert \Psi _{T}\rangle  \) is a vector and \( \hat{P} \) is
a matrix, then the procedure implied by \ref{gstate} is the algebraic
power method: If a matrix is applied iteratively to an initial arbitrary
vector for a sufficient number of times, only the dominant eigenvector,
\( \vert \Psi _{o}\rangle  \), will survive. One can see for large
\( i \), 

\begin{equation}
\label{gstate2}
\hat{P}^{i}\vert \Psi _{T}\rangle =\lambda ^{i}_{o}\langle \Psi _{o}\vert \Psi _{T}\rangle \vert \Psi _{o}\rangle +\mathcal{O}(\lambda _{1}^{i}),
\end{equation}

where \( \lambda _{o} \) is the leading eigenvalue, and \( \lambda _{1} \)
is the largest sub-leading eigenvalue.

For this approach, it is possible to obtain an estimator of the eigenvalue,
as described in \cite{JMH64}, given by,

\begin{equation}
\label{estimator eigenvalue}
\lambda _{o}=\lim _{i\to \infty }\left( \frac{\langle \phi \vert \hat{P}^{i+j}\vert \Psi _{T}\rangle }{\langle \phi \vert \hat{P}^{i}\vert \Psi _{T}\rangle }\right) ^{\frac{1}{m}}
\end{equation}

\subsubsection{Markov processes and stochastic projection}

\label{stochastic_projection_section}\index{Markov process}\index{stochastic projection}For
high dimensional vectors, such as those encountered in molecular electronic
structure, the algebraic power method described previously needs to
be generalized with stochastic implementation. For this to occur,
the projection operator must be symmetric, so that all eigenvalues
are real. This is the case for QMC methods, because \( \hat{P} \)
is a function of the Hamiltonian operator, \( \hat{H} \), which is
Hermitian by construction.

A stochastic matrix is a normalized non-negative matrix. By normalization,
we mean that the stochastic matrix columns add to one, \( \sum _{i}M_{ij}=1 \).
An \textbf{\( \mathbf{R} \)-}space representation would be a stochastic
propagator \( M(\mathbf{R},\mathbf{R}') \) that satisfies the condition

\begin{equation}
\label{normalized_propagator}
\int M(\mathbf{R},\mathbf{R}')d\mathbf{R}'=1.
\end{equation}

A Markov chain is a sequence of states obtained from subsequent transitions
from state \( i \) to \( j \) with a probability related to the
stochastic matrix element \( M_{ij} \), in which the move only depends
on the current state, \( i \). For example, in \( \mathbf{R} \)-space,
this is equivalent to the following process

\begin{eqnarray}
\pi (\mathbf{R}') & = & \int M(\mathbf{R}',\mathbf{R}'')\pi (\mathbf{R}'')d\mathbf{R}''\nonumber \\
\pi (\mathbf{R}) & = & \int M(\mathbf{R},\mathbf{R}')\pi (\mathbf{R}')d\mathbf{R}'\label{markov_chain_real_space} \\
 & \ldots  & \nonumber \label{markov_chain_2} 
\end{eqnarray}

The sequence of states \( S=\{\pi (\mathbf{R}''),\pi (\mathbf{R}'),\pi (\mathbf{R}),\ldots \} \)is
the Markov chain.

The propagators of QMC for electronic structure are not generally
normalized, therefore they are not stochastic matrices, but we can
represent them in terms of the latter by factoring, 

\begin{equation}
\label{PFactorization}
\hat{P}_{ij}=M_{ij}w_{j},
\end{equation}
 where the weights, \( w_{j} \), are defined by \( w_{j}=\sum _{i}\hat{P}_{ij} \).
This definition unambiguously defines both, the associated stochastic
matrix \( M \) and the weight vector \( w \). 

A MC sampling scheme of \( \hat{P}_{ij}\vert \Psi _{T}\rangle  \)
can be generated by first performing a random walk, and keep a weight
vector, \( W(\mathbf{R}) \) for the random walkers, 

{\footnotesize \begin{eqnarray}
\Psi (\mathbf{R}')\equiv \pi (\mathbf{R}')W(\mathbf{R}')=\int P(\mathbf{R}',\mathbf{R}'')\Psi (\mathbf{R}'')d\mathbf{R}'' & = & \int M(\mathbf{R}',\mathbf{R}'')B(\mathbf{R}')\Psi (\mathbf{R}'')d\mathbf{R}''\nonumber \\
\Psi (\mathbf{R})\equiv \pi (\mathbf{R})W(\mathbf{R})=\int P(\mathbf{R},\mathbf{R}')\Psi (\mathbf{R}')d\mathbf{R}' & = & \int M(\mathbf{R},\mathbf{R}')B(\mathbf{R})\Psi (\mathbf{R}')d\mathbf{R}'\label{branching_chain_real_space} \\
 & \ldots  & \nonumber 
\end{eqnarray}
}{\footnotesize \par}

Here, \index{branching}\( B(\mathbf{R}) \) is the function that
determines the weight of the configurations at each state of the random
chain. This leads to a generalized stochastic projection algorithm
for unnormalized transition probabilities that forms the basis for
population Monte Carlo (PopMC)\index{Monte Carlo!population algorithm}
algorithms, which are not only used for QMC, but they are also used
for statistical information processing and robotic vision \cite{YI}.
A generalized PopMC stochastic projection algorithm, represented in
\textbf{\( \mathbf{R} \)}-space follows,

\begin{enumerate}
\item \textbf{INITIALIZE}\\
Generate a set of \textbf{\( \mathbf{n} \)} random walkers, located
at different spatial positions, \( \mathcal{W}\equiv \{\mathbf{R}_{1},\mathbf{R}_{2},\ldots ,\mathbf{R}_{\mathbf{n}}\} \),
where \textbf{\( \mathbf{R}_{i} \)} denotes a Dirac Delta function
at that point in space, \( \delta (\mathbf{R}-\mathbf{R}_{i}) \).
These points are intended to sample a probability density function\index{probability density function (pdf)}
\( \Phi (\mathbf{R}) \). 
\item \textbf{MOVE} 

\begin{enumerate}
\item Each \index{random walker}walker \( j \) is moved independently
from \textbf{\( \mathbf{R} \)} to a new position \textbf{\( \mathbf{R}' \)},
according to the transition probability \\
\begin{equation}
\label{tprobpopmonte}
T(\mathbf{R}\to \mathbf{R}')\equiv M(\mathbf{R},\mathbf{R}')
\end{equation}

\item Ensure detailed balance if \( T(\mathbf{R}\to \mathbf{R}')\neq T(\mathbf{R}'\to \mathbf{R}) \),
by using a Metropolis\index{Monte Carlo!Metropolis} acceptance/rejection\index{acceptance probability}
step as in Eq. \ref{transition2}
\end{enumerate}
\item \textbf{WEIGHT} 

\begin{enumerate}
\item Calculate a weight vector\index{weigth vector} using a weighting
function \( B(\mathbf{R}_{i}) \)\begin{equation}
\label{reweight}
w^{*}_{i}=B(\mathbf{R}_{i})
\end{equation}
The ideal weight function preserves normalization of \( \hat{P}(\mathbf{R},\mathbf{R}') \)
and maintains individual weights, \( w_{i} \) close to unity. 
\item Update the weight of the walker, multiplying the weight of the previous
iteration by the weight of the new iteration,\\
\begin{equation}
\label{weight conservation}
w_{i}'=w^{*}_{i}*w_{i}
\end{equation}

\end{enumerate}
\item \textbf{RECONFIGURATION}

\begin{enumerate}
\item Split walkers with large weights into multiple walkers with weights
that add up to the original weight.
\item Remove walkers with small weight.
\end{enumerate}
\end{enumerate}
Step 4 is necessary to avoid statistical fluctuations in the weights.
It is a form of importance sampling in the sense that makes the calculation
stable over time. Some algorithms omit this step; see, for example,
efforts by \cite{dmcpuro:caffarel}, but it has been proved that such
calculations eventually diverge \cite{RAMC00}. There is a slight
bias associated with the introduction of step 4 together with population
control methods, that will be discussed in Sec. \ref{PopulationControl}.
When step 4 is used, \( B(\mathbf{R}) \) is also referred in the
literature as a \emph{branching factor}.

It is important to recall that PopMC algorithms are not canonical
Markov Chain Monte Carlo (MCMC) algorithms \cite{matamarillo, ADNdF+01},
in the sense that the propagator used is not normalized, and therefore
factoring the propagator into a normalized transition probability
and a weighting function is required.

\subsubsection{Projection operators or Green's functions}

Different projection operators lead to different QMC methods. If the
resolvent operator\index{resolvent operator},

\begin{equation}
\label{resolvent_gfmc}
\hat{P}(\hat{H})\equiv \frac{1}{1+\delta \tau (\hat{H}-E_{R})},
\end{equation}

is used, one obtains Green's function Monte Carlo (GFMC) \cite{MHK62, DMC86}.
This algorithm will be described in Sec. \ref{GFMCsection}

If the imaginary time evolution operator is used, i.e.,

\begin{equation}
\label{resolvent_dmc}
\hat{P}(\hat{H})\equiv e^{-\delta \tau (\hat{H}-E_{R})},
\end{equation}

one has the DMC method \cite{mc:qcrw, lesterfixednode}, which is
discussed in Sec. \ref{DMCSection}.

For finite \( \delta \tau  \), and for molecular systems, the exact
projector is not known analytically. In GFMC, the resolvent of Eq.
\ref{resolvent_gfmc} is sampled by iteration of a simpler resolvent,
whereas for DMC, the resolvent is known exactly at \( \tau \to 0 \)
, so an extrapolation to \( \delta \tau \to 0 \) is done.

Note that any decreasing function of \( \hat{H} \) can serve as a
projector. Therefore, new QMC methods still await to be explored.

\subsection{Imaginary propagator}

If one transforms the time-dependent Schr\"odinger equation (Eq. \ref{simpleschroeq})
to imaginary time\index{imaginary time}, i. e., 

\begin{equation}
\label{imaginary}
it\to \tau 
\end{equation}

then one obtains

\begin{equation}
\label{schroim}
\frac{\partial }{\partial \tau }\Psi (\mathbf{R},\tau )=(\hat{H}-E_{R})\Psi (\mathbf{R},\tau ).
\end{equation}

Here \( E_{R} \) is an energy offset, called the reference energy\index{reference energy}.
For real \( \Psi (\mathbf{R},\tau ) \), this equation has the advantage
of being an equation in \( \mathcal{R}^{N} \), whereas Eq. \ref{simpleschroeq},
has in general, complex solutions. 

Equation \ref{schroim} can be cast into integral form,

\begin{equation}
\label{schroint}
\Psi (\mathbf{R},\tau +\delta \tau )=\lambda _{\tau }\int G(\mathbf{R},\mathbf{R}',\delta \tau )\Psi (\mathbf{R}',\tau )d\mathbf{R}'
\end{equation}

The Green's function, \( G(\mathbf{R}^{'},\mathbf{R},\delta \tau ) \),
satisfies the same boundary conditions as Eq.. \ref{schroim},

\begin{equation}
\label{greeneq}
\frac{\partial }{\partial \tau }G(\mathbf{R},\mathbf{R}',\delta \tau )=(\hat{H}-E_{T})G(\mathbf{R},\mathbf{R}',\delta \tau )
\end{equation}

with the initial conditions associated with the propagation of a Dirac
delta function, namely,

\begin{equation}
\label{greenbc}
G(\mathbf{R},\mathbf{R}',0)=\delta (\mathbf{R}-\mathbf{R}')
\end{equation}

The form of the Green's function that satisfies Eq. \ref{greeneq},
subject to \ref{greenbc} is,

\begin{equation}
\label{propagator1}
G(\mathbf{R},\mathbf{R}',\delta \tau )=\langle \mathbf{R}\vert e^{-\tau (\hat{H}-E_{R})}\vert \mathbf{R}\rangle 
\end{equation}

This operator can be expanded in eigenfunctions, \( \Psi _{\alpha } \),
and eigenvalues \( E_{\alpha } \) of the system,

\begin{equation}
\label{basisrep}
G(\mathbf{R},\mathbf{R}',\delta \tau )=\sum _{\alpha }e^{-\tau (E_{\alpha }-E_{R})}\Psi ^{*}_{\alpha }(\mathbf{R}')\Psi _{\alpha }(\mathbf{R})
\end{equation}

For an arbitrary initial trial function, \( \Psi (\mathbf{R}) \),
in the long term limit, \( \tau \to \infty  \), one has

\begin{eqnarray}
\lim _{\tau \to \infty }e^{-\tau (\hat{H}-E_{T})}\Psi =\lim _{\tau \to \infty }\int G(\mathbf{R}',\mathbf{R},\tau )\Psi (\mathbf{R}')d\mathbf{R}'= &  & \label{limit} \\
\lim _{\tau \to \infty }\langle \Psi \vert \Psi _{o}\rangle e^{-\tau (E_{o}-E_{R})}\phi _{o}, & \nonumber 
\end{eqnarray}

and only the ground state wave function \( \Psi _{o} \) is obtained
from any initial wave function. Therefore, the imaginary time evolution
operator can be used as a projection operator as mentioned at the
beginning of this Chapter.

\subsubsection{Diffusion Monte Carlo stochastic projection}

\index{stochastic projection!diffusion Monte Carlo}Due to the high
dimensionality of molecular systems, a MC projection procedure is
used for obtaining expectation values. In this approach, the wave
function is represented as an ensemble of delta functions, also known
as configurations, walkers, or \emph{psips} (psi-particles)\index{random walker}\index{psi-particle}:

\begin{equation}
\label{deltas}
\Phi (R)\longleftrightarrow \sum _{k}\delta (\mathbf{R}-\mathbf{R}_{\mathbf{k}})
\end{equation}

The wave function is propagated in imaginary time using the Green's
function. In the continuous case, one can construct a Neumann series\index{Neumann series},

\begin{eqnarray}
\Psi ^{(2)}(\mathbf{R},\tau )=\lambda _{1}\int G(\mathbf{R}',\mathbf{R},\tau _{2}-\tau _{1})\Psi ^{(1)}(\mathbf{R}')d\mathbf{R} &  & \nonumber \\
\Psi ^{(3)}(\mathbf{R},\tau )=\lambda _{2}\int G(\mathbf{R}',\mathbf{R},\tau _{3}-\tau _{2})\Psi ^{(2)}(\mathbf{R}')d\mathbf{R} &  & \nonumber \\
(...) & \label{neumannnseries} 
\end{eqnarray}

This Neumann series is a specific case of the PopMC propagation of
Sec. \ref{stochastic_projection_section}. The discrete Neumann series
can be constructed in a similar way:

\begin{equation}
\label{discreteneumann}
\Phi ^{(n+1)}(\mathbf{R},\tau +\delta \tau )\longleftrightarrow \lambda _{k}\sum _{k}G^{(n)}(\mathbf{R},\mathbf{R}',\delta \tau )
\end{equation}

Therefore, a stochastic vector of configurations \( \mathcal{W}\equiv \{\mathbf{R}_{1},\ldots ,\mathbf{R}_{n}\} \)
is used to represent \( \Psi (\mathbf{R}) \) and is iterated using
\( G^{(n)}(\mathbf{R},\mathbf{R}',\delta \tau ) \).

\subsubsection{The form of the propagator}

\label{DMCSection}

\index{propagator}Sampling Eq. \ref{propagator1} can not be done
exactly, because the argument of the exponential is an operator composed
of two terms that do not commute with each other. 

In the short-time approximation \index{short-time approximation}(STA),
the propagator \( G(\mathbf{R},\mathbf{R}_{\mathbf{k}},d\tau ) \)
is approximated as if the kinetic and potential energy operators commuted
with each other,\\

\begin{eqnarray}
e^{(T+V)\delta \tau }\approx e^{T\delta \tau }\cdot e^{V\delta \tau }+\mathcal{O}((\delta \tau )^{2})\equiv G_{ST}\equiv  & G_{D}\cdot G_{B}\label{sta} 
\end{eqnarray}

The Green's function then becomes the product of a diffusion factor,
\( G_{D} \) and a branching\index{branching} factor \( G_{B} \).
Both propagators are known,

\begin{equation}
\label{gdiff}
G_{D}=(2\pi \tau )^{-3N/2}e^{-\frac{(\mathbf{R}-\mathbf{R}^{'})^{2}}{2\tau }},
\end{equation}
 and,

\begin{equation}
\label{gbranch}
G_{B}=e^{-\delta \tau (V(\mathbf{R})-2E_{T})}.
\end{equation}

\( G_{D} \) is a fundamental solution of the Fourier equation\index{diffusion!equation}
(that describes a diffusion process in wave function space) and \emph{\( G_{B} \)}
is the fundamental solution of a first-order kinetic birth-death process.

The Campbell-Baker-Haussdorff \index{Campbell-Baker-Haussdorff formula}(CBH)
formula,

\begin{equation}
\label{cbh}
e^{A}e^{B}=e^{A+B+\frac{1}{2}[A,B]+\frac{1}{12}[(A-B),[A,B]]+\ldots }
\end{equation}

can help in constructing more accurate decompositions, such as an
expansion with a cubic error \( \mathcal{O}((\delta \tau )^{3}) \),

\begin{equation}
\label{stathirdorder}
e^{\delta \tau (T+V)}=e^{\delta \tau (V/2)}e^{\delta \tau T}e^{\delta \tau (V/2)}+\mathcal{O}((\delta \tau )^{3})
\end{equation}

There are more sophisticated second order, \cite{SAC90}, and fourth
order, \cite{SAC97, HAF01} , expansions that reduce the error considerably
and make more exact DMC algorithms at the expense of a more complex
propagator. 

The most common implementation using \( G_{D} \) as a stochastic
transition probability \( T(\mathbf{R}\to \mathbf{R}') \), and \( G_{B} \)
as a weighting or branching factor, \( B(\mathbf{R}) \). Sampling
of Eq. \ref{gdiff} can be achieved by obtaining random variates from
a Gaussian distribution of standard deviation \( \delta _{\tau } \),
\( \mathcal{G}_{\delta \tau } \).

\subsection{Importance sampling}

\label{Importance Sampling}\index{importance sampling}

Direct application of the algorithm of the previous section to systems
governed by the Coloumb potential leads to large population fluctuations.
These arise because the potential \( \hat{V}(\mathbf{R}) \) becomes
unbounded and induces large fluctuations in the random walker population.
A remedy, importance sampling, was first used for GFMC in 1962 \cite{MHK62}
and extended to the DMC method in 1980 \cite{DMC80}.

In importance sampling, the goal is to reduce fluctuations, by multiplying
the probability distribution by a known trial function, \( \Psi _{T}(\mathbf{R}) \),
that is expected to be a good approximation for the wave function
of the system. Rather than \( \Psi (\mathbf{R},\tau ) \), one samples
the product,

\begin{equation}
\label{pairproduct}
f(\mathbf{R},\tau )=\Psi _{T}(\mathbf{R})\Psi (\mathbf{R},\tau )
\end{equation}

Multiplying Eq. \ref{schroint} by \( \Psi _{T}(\mathbf{R}) \), one
obtains,

\begin{equation}
\label{isgreen}
f(\mathbf{R},\tau +d\tau )=\int K(\mathbf{R}',\mathbf{R},\delta \tau )f(\mathbf{R}',\tau )d\mathbf{R}',
\end{equation}

where \( K(\mathbf{R},\mathbf{R}',\delta \tau )\equiv e^{-\tau (\hat{H}-E_{T})}\frac{\Psi _{T}(\mathbf{R})}{\Psi _{T}(\mathbf{R}')} \).
Expanding \( K \) in a Taylor series, at the \( \delta \tau \to 0 \)
limit, one obtains the expression,

\begin{equation}
\label{schmolowski_eq}
K=Ne^{-(\mathbf{R}_{2}-\mathbf{R}_{1}+\frac{1}{2}\nabla \ln \Psi _{T}(\mathbf{R}_{1})\delta \tau )^{2}/(2\delta \tau )}\times e^{-(\frac{\hat{H}\Psi _{T}(\mathbf{R}_{1})}{\Psi _{T}(\mathbf{R}_{1})}-E_{T})\delta \tau }\equiv K_{D}\times K_{B}
\end{equation}

Equation \ref{schmolowski_eq} is closely associated to the product
of the Kernel of the Smoluchowski equation\index{Smoluchowski equation},
which describes a diffusion process with drift, multiplied by a first
order rate process. Here the rate process is dominated by the local
energy, instead of the potential. The random walk is modified by appearance
of a drift term that moves configurations to regions of high values
of the wave function. This drift is the quantum force of Eq. \ref{quantum_force}.

The excess local energy\index{local energy!excess} \( (E_{T}-E_{L}(\mathbf{R})) \)
replaces the excess potential energy\index{potential energy!excess}
in the branching term exponent, see \ref{gbranch}. The local energy
has kinetic and potential energy contributions that tend to cancel
each other, giving a smoother function: If \( \Psi _{T}(\mathbf{R}) \)
is a reasonable function, the excess local energy will be nearly a
constant. The regions where charged particles meet have to be taken
care of by enforcing the cusp conditions on \( \Psi _{T}(\mathbf{R}) \)
(See Sec. \ref{PropertiesExact}).

The local energy\index{local energy!estimator} is the estimator of
the energy with a lower statistical variance, so it is preferred over
other possible choices for an estimator. A simple average of the local
energy will yield the estimator of the energy of the quantum system%
\footnote{For other energy estimators, refer to the discussion in \cite{DMC86}
and \cite{azul}.
},

\begin{eqnarray}
\langle E_{L}\rangle  & = & \int f(\mathbf{R},\tau \to \infty )E_{L}(\mathbf{R})d\mathbf{R}/\int f(\mathbf{R})d\mathbf{R}\label{local_energy_estimator} \\
 & = & \int \Psi (\mathbf{R})\Psi _{T}(\mathbf{R})\left[ \frac{\hat{H}\Psi _{T}(\mathbf{R})}{\Psi _{T}(\mathbf{R})}\right] d\mathbf{R}/\int \Psi (\mathbf{R})\Psi _{T}(\mathbf{R})\nonumber \\
 & = & \int \Psi (\mathbf{R})\hat{H}\Psi (\mathbf{R})d\mathbf{R}/\int \Psi (\mathbf{R})\Psi _{T}(\mathbf{R})d\mathbf{R}\nonumber \\
 & = & E_{o}\nonumber 
\end{eqnarray}

Therefore, a simple averaging of the local energy, will yield the
DMC energy estimator:

\begin{equation}
\label{estimator_elocal_Sum}
\langle E_{L}\rangle =\lim _{N_{s}\to \infty }\frac{1}{N_{s}}\sum _{i}^{N_{w}}E_{L}(\mathbf{R}_{i})
\end{equation}

Because the importance sampled propagator, \( K(\mathbf{R},\mathbf{R}',\delta \tau ) \),
is only exact to a certain order, for obtaining an exact estimator
is necessary to extrapolate\index{time step!extrapolation} to \( \delta \tau =0 \)
for several values of \( \langle E_{L}\rangle  \).

Importance sampling with appropriate trial functions, such as those
used for accurate VMC calculations, can increase the efficiency of
the random walk by several orders of magnitude. In the limit required
to obtain the exact trial function, only a single evaluation of the
local energy is required to obtain the exact answer. Importance sampling
has made molecular and atomic calculations feasible. Note that the
quantum force present in Eq. \ref{schmolowski_eq} also moves random
walkers away from the nodal regions into regions of large values of
the trial wave function, reducing the number of attempted node crossings
by several orders of magnitude\index{fermion nodes!crossings}.

\subsection{Population control}

\index{population control}\label{PopulationControl}

If left uncontrolled, the population of random walkers\index{random walker!ensemble}
will eventually vanish or fill all computer memory. Therefore, some
form of population control is needed to stabilize the number of random
walkers. Control is usually achieved by slowly changing \( E_{T} \)
as the simulation progresses. As more walkers are produced in the
procedure, one needs to lower the trial energy, \( E_{T} \), or if
the population starts to decrease, then one needs to raise \( E_{T} \).
This can be achieved by periodically changing the trial energy. One
version of the adjustment is to use,

\begin{equation}
\label{et_adjust}
E_{T}=\langle E_{o}\rangle +\alpha \ln \frac{N^{0}_{w}}{N_{w}},
\end{equation}

where \( <E_{o}> \) is the best approximation to the eigenvalue of
the problem to this point, \( \alpha  \) is a parameter that should
be as small as possible while still having a population control effect,
\( N^{0}_{w} \) is the number of desired random walkers, and \( N_{w} \)
is the current number of random walkers.

This simple population control procedure has a slight bias if the
population control parameter \( \alpha  \) is large, or if the population
is small. The bias observed goes as \( 1/N_{w} \), and, formally
a \( N_{w}\to \infty  \) extrapolation is required\index{finite population bias}.
The bias is absent in the limit of an infinite population.

A recently resurrected population control strategy, stochastic reconfiguration
\cite{SS98, MCB98, SSLC00, RAMC00} originally came from a the work
of \cite{JHH84}. In this algorithm, walkers carry a weight, but the
weight is redetermined at each step to keep the population constant.
The idea behind this method is to control the global weight \( \bar{w} \)
of the population,

\begin{equation}
\label{global_weight}
\bar{w}=\frac{1}{N_{w}}\sum ^{N_{w}}_{i=1}w_{i}
\end{equation}

The idea is to introduce a renormalized individual walker weight,
\( \omega _{i} \), defined as,

\begin{equation}
\label{renormalized_weight}
\omega _{i}\equiv \frac{w_{i}}{\bar{w}}.
\end{equation}

Another stochastic reconfiguration\index{stochastic reconfiguration}
scheme proposes setting the number of copies of walker \( i \) for
the next step, proportional to the renormalized walker weight, \( \omega _{i} \).
This algorithm has shown to have less bias than the scheme of Eq.
\ref{et_adjust}, and also has the advantage of having the same number
of walkers at each step, simplifying implementations of the algorithm
in parallel computers.

\subsection{Diffusion Monte Carlo algorithm}

\textbf{\label{DMCSectionAlgorithm}\index{Monte Carlo!diffusion}}

There are several versions of the DMC algorithm. The approach presented
here focuses on simplicity. For the latest developments, the reader
is referred to the references \cite{lesterfixednode, MFD88, diffsmallerror}.

\begin{enumerate}
\item Initialize an ensemble \( \mathcal{W} \) of \( N_{\mathcal{W}} \)
configurations, distributed according to \( P(\mathbf{R}) \) for
\( \Psi _{T}(\mathbf{R}) \); for example, use the random walkers
obtained from a previous VMC run.
\item For every configuration in \( \mathcal{W} \),

\begin{enumerate}
\item Propose an electron move from \( \Psi (\mathbf{R})\equiv \Psi (\mathbf{r}_{1},\mathbf{r}_{2},\ldots ,\mathbf{r}_{i},\ldots ,\mathbf{r}_{N_{p}}) \)
to \( \Psi (\mathbf{R}')\equiv \Psi (\mathbf{r}_{1},\mathbf{r}_{2},\ldots ,\mathbf{r}'_{i},\ldots ,\mathbf{r}_{N_{p}}) \).
The short-time approximation propagator, \( K(\mathbf{R},\mathbf{R}';\delta \tau ) \),
has an associated stochastic move, \\
\begin{equation}
\label{dmc_move}
\mathbf{R}'\to \mathbf{R}+\mathbf{F}_{\mathbf{q}}(\mathbf{R})\delta \tau +\mathcal{G}_{\delta \tau }
\end{equation}

\item Enforce the fixed node constraint: if a random walker crosses a node,
i.e. \( \textrm{sign}(\Psi _{T}(\mathbf{R}))\neq \textrm{sign}(\Psi _{T}(\mathbf{R}')) \),
then reject the move for the current electron and proceed to treat
the next electron.
\item Compute the Metropolis\index{Monte Carlo!Metropolis} acceptance/rejection
probability\\
\begin{equation}
\label{transition3}
P(\mathbf{R}\to \mathbf{R}')=\min \left( 1,\frac{K_{D}(\mathbf{R},\mathbf{R}';\delta \tau )\Psi _{T}^{2}(\mathbf{R}')}{K_{D}(\mathbf{R}',\mathbf{R};\delta \tau )\Psi ^{2}_{T}(\mathbf{R})}\right) ,
\end{equation}
where \( K_{D} \) is the diffusion and drift transition probability
given by Eq. \ref{schmolowski_eq}. 
\item Compare \( P(\mathbf{R}\to \mathbf{R}') \) with an uniform random
number between 0 and 1, \( \mathcal{U}_{[0,1]} \), if \( P>\mathcal{U}_{[0,1]} \),
accept the move, otherwise, reject it.
\end{enumerate}
\item Calculate the branching\index{branching} factor \( G_{B} \) for
the current configuration\\
\begin{equation}
\label{branching_dmc_algorithm}
B(\mathbf{R},\mathbf{R}')=e^{(E_{R}-\frac{1}{2}(\frac{\hat{H}\Psi _{T}(\mathbf{R}')}{\Psi _{T}(\mathbf{R}')}+\frac{\hat{H}\Psi _{T}(\mathbf{R})}{\Psi _{T}(\mathbf{R})}))\delta \tau }.
\end{equation}

\item Accumulate all observables, such as the energy. All contributions,
\( O_{i} \), are weighted by the branching factor, i. e., \\
\begin{equation}
\label{accumulation}
O^{(n+1)}_{T}=O^{(n)}_{T}+B(\mathbf{R},\mathbf{R}')O_{i}(\mathbf{R}),
\end{equation}
where \( O_{T}^{(n)} \) is the cumulative sum of the observable at
step \( n \).
\item Generate a new generation of random walkers, reproducing the existing
population, creating an average \( B(\mathbf{R},\mathbf{R}') \) new
walkers out of a walker at \( \mathbf{R} \). The simplest procedure
for achieving this goal is to generate \( n \) new copies of \( \mathbf{R} \)
where \( n=\textrm{int}(B(\mathbf{R},\mathbf{R}')+\mathcal{U}_{[0,1]}) \).
\item Perform blocking statistics\index{blocking statistics} (see Sec.
\ref{Statistics}), and apply population control\index{population control}
(see Sec. \ref{PopulationControl})

\begin{enumerate}
\item One choice is to update the reference energy, \( E_{R} \) at the
end of each accumulation block, \\
\begin{equation}
\label{simple_dmc_update}
E_{R}\leftarrow E_{R}+E^{\omega }_{R}*E_{B},
\end{equation}
 where \( E^{\omega }_{R} \) is a re-weighting parameter, usually
chosen to be \( \approx 0.5 \), and \( E_{B} \) is the average energy
for block \( B \), \( E_{B}=E^{sum}/N_{B} \)
\item Discard a relaxation time of steps, \( N_{Rel} \), which is of the
order of a tenth of a block, because moving the reference energy induces
the most bias in about one relaxation time.
\end{enumerate}
\item Continue the loop until the desired accuracy is achieved.
\end{enumerate}
Umrigar \emph{et al.} \cite{diffsmallerror} proposed several modifications
to the above algorithm to reduce time-step error. These modifications
concentrate on improving the propagator in regions where the short-time
approximation performs poorly; namely, near wave function nodes and
Coulomb singularities\index{wave function!singularities}. These propagator
errors are expected, because the short-time approximation propagator
assumes a constant potential over the move interval, which is a poor
approximation in \( \Psi  \) regions where the Coulomb interaction
diverges\index{Coulomb interactions}.

\subsection{Green's function Monte Carlo}

\textbf{\index{Monte Carlo!Green's function}\label{GFMCsection}}

The GFMC method is a QMC approach that has the advantage of having
no time-step error\index{time step!error}. It has been shown to require
more computer time than DMC, and therefore, has been applied less
frequently than DMC to atomic and molecular systems. Good descriptions
of the method can be found in \cite{MHK62, DMC83, mosko:coloumb, kalosmontecarlo, KES86b}.
The GFMC approach is a PopMC method for which the projector for obtaining
the ground state Green's function is the standard resolvent for the
Schrödinger equation\index{resolvent operator } (see Eq. \ref{resolvent_gfmc}).
The integral equation for this case, takes the simple form,

\begin{equation}
\label{iteration_gfmc}
\Psi ^{(n+1)}=\left[ \frac{E_{T}+E_{C}}{\hat{H}+E_{C}}\right] \Psi ^{(n)}
\end{equation}

where the constant \( E_{C} \) is positive and fulfills the condition
that \( \vert E_{C}\vert >\vert E_{o}\vert  \), and \( E_{T} \)
is a trial energy. The resolvent of Eq. \ref{iteration_gfmc} is related
to the DMC propagator by the one-sided Laplace transform,

\begin{equation}
\label{mc_integral_gfmc}
\frac{1}{\hat{H}+E_{C}}=\int ^{\infty }_{0}e^{-(\hat{H}+E_{C})\tau }d\tau 
\end{equation}

This integral is evaluated by MC. After equilibration, the sampled
times have a Poisson distribution\index{Poisson distribution} with
a mean of \( \frac{N_{s}}{E_{o}+E_{C}} \) after \( N_{s} \) steps.
\( E_{C} \) is a parameter that controls the average time step.

The Green's function is not known in close form, so it has to be sampled
by MC. This can be done by rewriting the resolvent in the form,

\begin{equation}
\label{huhcexpansiongfmc}
\frac{1}{\hat{H}+E_{C}}=\frac{1}{\hat{H}_{U}+E_{C}}+\frac{1}{\hat{H}_{U}+E_{C}}(\hat{H}_{U}-\hat{H})\frac{1}{\hat{H}+E_{C}}
\end{equation}

The Hamiltonian \( \hat{H}_{U} \) represents a family of solvable
Hamiltonians. To sample the Green's function, one samples the sum
of terms on the right-hand side of Eq. \ref{huhcexpansiongfmc}..
The Green's functions associated with \( \hat{H} \) and \( \hat{H}_{U} \),
satisfy the relations,

\begin{equation}
\label{eqforgf1}
(\hat{H}+E_{C})G(\mathbf{R},\mathbf{R}')=\delta (\mathbf{R}-\mathbf{R}')
\end{equation}

\begin{equation}
\label{eqforgf2}
(\hat{H}_{U}+E_{C})G_{U}(\mathbf{R},\mathbf{R}')=\delta (\mathbf{R}-\mathbf{R}')
\end{equation}

The most commonly used form of \( \hat{H}_{U} \) is,

\begin{equation}
\label{simple_domain_hamiltonian}
\hat{H}_{U}=\frac{1}{2}\nabla _{R}^{2}+U
\end{equation}

where \( U \) is a potential that is independent of \( \mathbf{R} \).
It is convenient to have \( G_{U}(\mathbf{R},\mathbf{R}') \) vanish
at the domain boundary. \( \hat{H}_{U} \) should be a good approximation
to \( \hat{H} \) in the domain to achieve good convergence.

The \( \mathbf{R} \)-space representation of Eq. \ref{huhcexpansiongfmc}
is

\begin{eqnarray}
G(\mathbf{R},\mathbf{R}')=G_{U}(\mathbf{R},\mathbf{R}')-\int _{S}d\mathbf{R}''G(\mathbf{R},\mathbf{R}'')\left[ -\hat{n}\cdot \nabla G_{U}(\mathbf{R}'',\mathbf{R})\right]  &  & \nonumber \\
+\int _{V}d\mathbf{R}''G(\mathbf{R},\mathbf{R}'')\left[ U-V(\mathbf{R}'')\right] G_{U}(\mathbf{R}'',\mathbf{R}') & \label{greensf} 
\end{eqnarray}

\subsection{Fixed-node approximation}

\label{Section:fixednode}

\index{fermion nodes}We have not discussed the implications of the
fermion character of \( \Psi (\mathbf{R}) \). It is an excited state
in a manifold containing all the fermionic and bosonic states. A fermion
wave function has positive and negative regions that are difficult
to sample with the DMC algorithm as described in Sec. \ref{DMCSectionAlgorithm}.
Considering real wave functions, \( \Psi (\mathbf{R}) \) contains
positive and negative regions, \( \Psi ^{+}(\mathbf{R}) \), and \( \Psi ^{-}(\mathbf{R}) \)
that, in principle, could be represented as probabilities. The sign
of the wave function could be used as an extra weight for the random
walk. In practice, this is a very slowly convergent method.

Returning to the importance sampled algorithm, recall that the initial
distribution, \( \vert \Psi (\mathbf{R})\vert ^{2} \), is positive
. Nevertheless, the Green's function, \( K(\mathbf{R},\mathbf{R}') \),
can become negative, if a random walker crosses a node of the trial
wave function. Again, the sign of \( K(\mathbf{R},\mathbf{R}') \)
could be used as a weight for sampling \( \vert K(\mathbf{R},\mathbf{R}')\vert  \).
The problem is that the statistics of this process lead to exponential
growth of the variance of any observable.

The simplest approach to avoid exponential growth is to forbid moves
in which the product wave function, \( \Psi (\mathbf{R})\Psi _{T}(\mathbf{R}) \),
changes sign. This boundary condition on permitted moves is the defining
characteristic of the fixed-node approximation\index{fixed-node approximation (FNA)}
(FNA). The nodes of the sampled wave function are \emph{fixed} to
be the nodes of the trial wave function. The FNA is an inherent feature
of the DMC method, which is, by far, the most commonly used method
for atomic and molecular MC applications \cite{mc:qcrw, lesterfixednode}.

The fixed-node energy is an upper bound to the exact energy of the
system. In fact, it is the best solution for that fixed set of nodes.
The DMC method has much higher accuracy than the VMC method. For atomic
and molecular systems, it is common to recover \( 95-100\% \) of
the CE, cf Sec. \ref{NumericalSolution}, whereas the CE recovered
with the VMC approach is typically less than \( 80\% \) of the total. 

\subsection{Exact methods}

\index{Monte Carlo!exact}Probably the most important algorithmic
challenge that still remains to be explored is the {}``node problem''.
Although progress has been made in systems that contain up to a dozen
of electrons \cite{schmidt:gfmc, DCBJA84, SZMHK91, JBA91, ZLSZ94},
a stable algorithm that can sample the exact wave function without
resorting to the FNA remains to be determined. In this section, we
discuss a family of methods that avoid the FNA. These approaches yield
exact answers, usually associated with a large increase in computational
time.

The Pauli antisymmetry\index{antisymmetry} principle imposes a boundary
condition on the wave function. It is the requirement that the exchange
of like-spin electrons changes the sign of the wave function. This
condition is a global condition that has to be enforced within an
algorithm that only considers individual random walkers, i.e., a local
algorithm. The FNA is the most commonly imposed boundary condition.
It satisfies the variational principle, i.e, FN solutions approach
the exact energy from above. This is an useful property, but one that
does not assist the search for exact results, because there is not
an easy way to parametrize the nodal surface and vary it to obtain
the exact solution. We now describe methods that impose no additional
boundary conditions on the wave function.

\subsubsection{The release node method}

\label{RNMethod}

\index{released node method}The evolution operator, \( e^{-\tau (\hat{H}-E_{T})} \),
is symmetric and has the same form for both fermions and bosons. Straightforward
application of it to an arbitrary initial wave function, \( \vert \Psi _{o}\rangle  \),
leads to collapse to the ground state bosonic wave function, as can
be seen from Eq. \ref{limit}. 

An arbitrary fermion wave function, \( \Psi (\mathbf{R}) \), can
be separated into two functions, \( \Psi ^{+}(\mathbf{R}) \), and
\( \Psi ^{-}(\mathbf{R}) \), as follows,

\begin{equation}
\label{psiplusminus}
\Psi ^{\pm }(\mathbf{R},\tau )\equiv \frac{1}{2}\left[ \vert \Psi (\mathbf{R},\tau )\vert \pm \Psi (\mathbf{R},\tau )\right] .
\end{equation}

Note that the original trial wave function is recovered as,

\begin{equation}
\label{recoveringpsi}
\Psi (\mathbf{R},\tau )=\Psi ^{+}(\mathbf{R},\tau )-\Psi ^{+}(\mathbf{R},\tau )
\end{equation}

The released node (RN) algorithm involves two independent DMC calculations,
using \( \Psi ^{+} \) and \( \Psi ^{-} \) as the wave functions
to evolve,

\begin{eqnarray}
\Psi (\mathbf{R},\tau )=\int G(\mathbf{R},\mathbf{R}',\delta \tau )\Psi (\mathbf{R}',0)d\mathbf{R}' & = & \label{released_evolve} \\
\int G(\mathbf{R},\mathbf{R}',\tau )\Psi ^{+}(\mathbf{R}',0)d\mathbf{R}'-\int G(\mathbf{R},\mathbf{R}',\tau )\Psi ^{-}(\mathbf{R}',0)d\mathbf{R}' & = & \nonumber \\
\Psi ^{+}(\mathbf{R},\tau )-\Psi ^{-}(\mathbf{R},\tau ) & \nonumber 
\end{eqnarray}

The time evolution of the system can be followed from the difference
of separate simulations for \( \Psi ^{\pm }(\mathbf{R}) \). Note
that both distributions are always positive during the simulation,
and that they decay to the ground state bosonic wave function. This
decay is problematic because the {}``signal-to-noise'' ratio in
this method depends on the difference between these two distributions.
The decay of the difference \( \Psi ^{+}(\mathbf{R},\tau )-\Psi ^{-}(\mathbf{R},\tau ) \)
goes roughly as \( e^{\tau (E_{F}-E_{B})} \), where \( E_{F} \)
is the lowest fermion state energy and \( E_{B} \) is the bosonic
ground state energy. 

For this method to be practical, one needs to start with the distribution
of a good fermion trial wave function. The distribution will evolve
from this starting point to the bosonic ground state at large imaginary
time \( \tau  \). In an intermediate {}``transient'' regime one
can collect information on the exact fermion wave function\index{transient estimator}.

The energy can be estimated from the expression,

\begin{eqnarray}
E_{RN}(\tau ) & = & \frac{\int \Psi (\mathbf{R},\tau )\hat{H}\Psi _{T}(\mathbf{R})d\mathbf{R}}{\int \Psi (\mathbf{R},\tau )\Psi _{T}d\mathbf{R}}\label{transient_estimator} \\
 & = & \frac{\int \Psi ^{+}(\mathbf{R},\tau )\hat{H}\Psi _{T}(\mathbf{R})d\mathbf{R}}{\int \left[ \Psi ^{+}(\mathbf{R},\tau )-\Psi ^{-}(\mathbf{R},\tau )\right] \Psi _{T}(\mathbf{R})d\mathbf{R}}-\frac{\int \Psi ^{-}(\mathbf{R},\tau )\hat{H}\Psi _{T}(\mathbf{R})d\mathbf{R}}{\int \left[ \Psi ^{+}(\mathbf{R},\tau )-\Psi ^{-}(\mathbf{R},\tau )\right] \Psi _{T}(\mathbf{R})d\mathbf{R}}\nonumber \label{transient_estimator} \\
 & = & E_{F}\nonumber 
\end{eqnarray}

In the release node method \cite{DCBJA84}, a fixed-node distribution
is propagated as usual, but now two sets of random walkers are retained,
\( \mathcal{W}_{FN} \), the fixed node ensemble and \( \mathcal{W}_{RN} \),
the released node ensemble. Walkers are allowed to cross nodes, and
when they do, they are transferred from \( \mathcal{W}_{FN} \) to
\( \mathcal{W}_{RN} \). Also a account is made of the number of iterations
that a walker has survived, \( \mathcal{S}_{RN}=\{s_{1},\ldots ,s_{N_{w}}\} \).
This index is used to bin the walkers by age. Each time a walker crosses
a node, a summation weight associated with it, \( \Omega _{RN}=\{\omega _{1},\ldots ,\omega _{N_{w}}\} \)
changes sign. These weights determine the sign of the walker contribution
to global averages.

The released node energy can be calculated using the estimator,

\begin{equation}
\label{released_node_estimator}
E_{RN}=\frac{\sum ^{Nw}_{i=1}\omega _{i}\frac{\Psi _{T}(\mathbf{R}_{i})}{\Psi (\mathbf{R}_{i})}E_{L}(\mathbf{R}_{i})}{\sum ^{Nw}_{i=1}\omega _{i}\frac{\Psi _{T}(\mathbf{R}_{i})}{\Psi (\mathbf{R}_{i})}}.
\end{equation}

\subsubsection{Fermion Monte Carlo}

\label{FMCSection}

\index{Monte Carlo!fermion}From the previous section one can infer
that if a method in which the distribution does not go to the bosonic
ground state, but stays in an intermediate regime, will not have the
deficiency of exponential growth of {}``signal to noise''. This
leads to the fermion Monte Carlo (FMC) method. The approach \cite{MHK97, MHK98B, PK99, MHK00, MHK00b}
involves correlated random walks that achieve a constant {}``signal
to noise''. 

The expectation value of Eq. \ref{released_node_estimator} for an
arbitrary distribution of signed walkers can be rewritten as

\begin{equation}
\label{E_FMC}
\langle E_{FMC}\rangle =\frac{\sum ^{Nw}_{i=1}[\frac{\hat{H}\psi _{T}(\mathbf{R}_{i}^{+})}{\Psi ^{+}_{G}(R_{i}^{+})}-\frac{\hat{H}\psi _{T}(\mathbf{R}_{i}^{-})}{\Psi ^{-}_{G}(R_{i}^{-})}}{\sum ^{Nw}_{i=1}[\frac{\psi _{T}(R_{i}^{+})}{\Psi ^{+}_{G}(R_{i}^{+})}\frac{\psi _{T}(R_{i}^{-})}{\Psi ^{-}_{G}(R_{i}^{-})}]}
\end{equation}

where \( \Psi ^{\pm }_{G}(\mathbf{R}^{\pm }) \) are the guiding functions
for a pair of random walkers \( P_{i}=\{\mathbf{R}_{i}^{+},\mathbf{R}^{-}_{i}\} \).
Note that the variance of the energy estimator of Eq. \ref{E_FMC}
goes to infinity as the difference between the two populations goes
to zero, i. e. the denominator,

\begin{equation}
\label{denominator}
\mathcal{D}\equiv \sum ^{Nw}_{i=1}[\frac{\psi _{T}(R_{i}^{+})}{\Psi ^{+}_{G}(R_{i}^{+})}\frac{\psi _{T}(R_{i}^{-})}{\Psi ^{-}_{G}(R_{i}^{-})}],
\end{equation}

goes to zero as the simulation approaches to the bosonic ground state.
A procedure that would not change \( \langle E_{FMC}\rangle  \) would
be to cancel positive and negative random walkers whenever they meet
\cite{schmidt:gfmc}. Although random walks are guaranteed to meet
in one dimension, they need not meet in several dimensions, due to
the exponentially decaying walker density in \( \mathbf{R} \)-space.
Besides, cancellation has to be combined with other procedures to
insure a stable algorithm.

\index{random walker!cancellation}Cancellation can be increased by
introducing correlation between the random walkers. Recall the diffusion
step in DMC, in which walkers diffuse from \( \mathbf{R} \) to \( \mathbf{R}' \)
following \( G_{D} \) of Eq. \ref{gdiff}. In the DMC algorithm,
this is implemented stochastically by updating the coordinates of
the random walkers with a random displacement taken from a Gaussian
distribution with a variance of \( \delta \tau  \), 

\begin{equation}
\label{fmc_gaussian_move}
\mathbf{R}'^{+}\to \mathbf{R}^{+}+\mathcal{G}^{+}_{\delta \tau }\textrm{ and }\mathbf{R}^{-}\to \mathbf{R}^{-}+\mathcal{G}^{-}_{\delta \tau }.
\end{equation}

If we introduce correlation between the Gaussian vectors, \( \mathcal{G}^{+}_{\delta \tau } \)
and \( \mathcal{G}^{-}_{\delta \tau } \), the expectation value of
Eq. \ref{E_FMC} is not affected, because it is linear in the density
of random walkers\index{diffusion!correlated}.

An efficient cancellation scheme can be achieved if the Gaussian vectors
are correlated as follows,

\begin{equation}
\label{reflected_dynamics}
\mathcal{G}^{-}_{\delta \tau }=\mathcal{G}^{+}_{\delta \tau }U^{-}-2(\mathcal{G}^{+}_{\delta \tau }\cdot \frac{(\mathbf{R}^{+}-\mathbf{R}^{-})}{\vert \mathbf{R}^{+}-\mathbf{R}^{-}\vert ^{2}})\cdot (\mathbf{R}^{+}-\mathbf{R}^{-}),
\end{equation}
Equation \ref{reflected_dynamics}\index{dynamics!fictitious!correlated}
accounts for reflection along the perpendicular bisector of the vector
that connects the pair, \textbf{\( \mathbf{R}^{+}-\mathbf{R}^{-} \)}.
This cancellation scheme generates a correlated random walk in one
dimension along the vector \( \mathbf{R}^{+}-\mathbf{R}^{-} \). This
one-dimensional random walk is independent of the number of dimensions
of the physical system, and therefore, overcomes the cancellation
difficulties mentioned above. Walkers are guaranteed to meet under
these conditions.

The modifications to the DMC algorithm mentioned to this point are
necessary, but not sufficient for achieving a stable algorithm. If
one were to interchange the random walker populations, \( \{\mathbf{R}^{+}_{1},\ldots ,\mathbf{R}^{+}_{N_{w}}\}\leftrightarrow \{\mathbf{R}^{-}_{1},\ldots ,\mathbf{R}^{-}_{N_{w}}\} \),
the fictitious dynamics would not be able to distinguish between the
two populations, leading to a random walk with two degenerate ground
states. Namely, a ground state in which all the positive walkers,
\( \mathbf{R}^{+} \) are marginally on the positive region of the
wave function, and vice versa, \( \{\Psi ^{+}(\mathbf{R}^{+}),\Psi ^{-}(\mathbf{R}^{-})\} \)
and \( \{\Psi ^{+}\{\mathbf{R}^{-}\},\Psi ^{-}(\mathbf{R}^{+})\} \).
This \emph{plus-minus} symmetry can be broken by using two distinct
guiding functions. For example, the guiding function

{\small \begin{equation}
\label{guiding_functions}
\Psi ^{\pm }_{G}=\sqrt{\Psi ^{2}_{S}(\mathbf{R})+c^{2}\Psi ^{2}_{A}(\mathbf{R})}\pm c\Psi _{A}(\mathbf{R}),
\end{equation}
}{\small \par}

where \( \Psi _{S}(\mathbf{R}) \) is a symmetric function under permutation
of electron labels; \( \Psi _{A}(\mathbf{R}) \) is an antisymmetric
function, and \( c \) is a small adjustable parameter. The guiding
functions of Eq. \ref{guiding_functions} are almost equal, which
provides nearly identical branching factors for the walker pair. It
is positive everywhere, a requirement for the DMC algorithm, and it
is symmetric under permutation of the coordinates, {\small \( \Psi ^{+}_{G}(\hat{\mathbf{P}}\mathbf{R})=\Psi ^{-}_{G}(\mathbf{R}) \).}{\small \par}

The use of different guiding functions is the last required ingredient
for a stable algorithm. It breaks the \emph{plus-minus} symmetry effectively,
because the drift dynamics is different because the quantum force
of Eq. \ref{quantum_force} is distinct for each population.

For a complete description of the FMC algorithm, the reader is referred
to \cite{MHK00}.

The denominator of Eq. \ref{denominator} is an indicator of stability
of the algorithm. It is a measure of the antisymmetric component of
the wave function. FMC calculations have shown stable denominators
for thousands of relaxation times, indicating the stability of the
fermion algorithm.

Early versions of the method \cite{schmidt:gfmc} do not scale well
with system size, due to the use of uncorrelated cancellation schemes.
Nevertheless, researchers have been applied successfully to several
small molecular systems obtaining solutions to the Schrödinger equation
with no systematic error \cite{SZMHK91, ABJBA94, ABJBA94-a, JBA01}.
This version of the FMC algorithm, with GFMC propagation and without
correlated dynamics, is known as exact quantum Monte Carlo (EQMC).

\subsection{Zero variance principle}

\label{ZeroVariance}\index{zero-variance principle}An increase in
computational efficiency can be achieved by improving the observables
\( \hat{O} \) by renormalizing them to observables that have the
same expectation value, but lower variance. Recent work \cite{RAMC99, RAMC00b}
has shown that estimators for the energy and energy derivatives with
respect to nuclear coordinates can be constructed.

One can propose a trial operator \( \hat{H}_{V} \) and auxiliary
trial function\index{trial function!auxiliary}, \( \Psi _{V} \)
such that the evaluation of a renormalized observable \( \overline{O} \)
will have a variance that is smaller than that of the original observable
\( \hat{O} \), and in principle can even be suppressed.

To develop this concept, let us construct a trial operator \index{trial operator}\( \hat{H}_{V} \)
such that,

\begin{equation}
\label{prop1}
\int \hat{H}_{V}(\mathbf{R},\mathbf{R}')\sqrt{\pi (\mathbf{R}')}d\mathbf{R}'=0,
\end{equation}

where \( \pi (\mathbf{R}') \) is the MC distribution. For example,
in VMC the MC distribution is the wave function squared, \( \Psi _{T}(\mathbf{R})^{2} \),
and in DMC it is the mixed distribution of Eq. \ref{pairproduct}.
Next, propose a renormalized observable \( \overline{O}(\mathbf{R}) \)
related to the observable \( \hat{O}(\mathbf{R}) \) given by,

\begin{equation}
\label{feqn1}
\overline{O}(\mathbf{R})=\hat{O}(\mathbf{R})+\frac{\int \hat{H}_{V}(\mathbf{R},\mathbf{R}')\psi _{V}(\mathbf{R}')d\mathbf{R}'}{\sqrt{\pi (\mathbf{R}')}}
\end{equation}

The mean of the rescaled operator is formally, 

\begin{equation}
\label{obs_avg}
<\overline{O}>=\frac{\int \hat{O}(\mathbf{R})\pi (\mathbf{R})d\mathbf{R}+\frac{\int \int \pi (\mathbf{R})\hat{H}_{V}(\mathbf{R},\mathbf{R}')\Psi _{V}(\mathbf{R}')d\mathbf{R}d\mathbf{R}'}{\sqrt{\pi (\mathbf{R})}}}{\int \pi (\mathbf{R})d\mathbf{R}}
\end{equation}

which, by property \ref{prop1}, is the same as the mean for the unnormalized
operator:

\begin{equation}
\label{samemeans}
<\overline{O}>=<\hat{O}>
\end{equation}

Operator \( \overline{O} \) can be used as an unbiased estimator,
even though statistical errors for \( \overline{O} \) and \( \hat{O} \)
can be quite different. The goal of this kind of importance sampling
is to reduce the fluctuations by construction of such an operator.

The implementation of the procedure requires the optimization of a
set of parameters for the auxiliary trial wave function, \( \Psi _{V}(\mathbf{R},\Lambda _{V}) \),
using the minimization functional,

\begin{equation}
\label{zerovar_functional}
\int \hat{H}_{V}(\mathbf{R},\mathbf{R}')\Psi _{V}(\mathbf{R},\Lambda _{V})d\mathbf{R}'=-[\bar{O}(x)-<\bar{O}>]\sqrt{\pi (\mathbf{R})}
\end{equation}

After the parameters \( \Lambda _{V} \) are optimized, one can run
a simulation to average \( \overline{O}, \) instead of \( \hat{O} \).
The choice of auxiliary Hamiltonian suggested by recent work \cite{RAMC99}
is

\begin{equation}
\label{auxilary_hamiltonian}
H_{V}(x,y)=-\frac{1}{2}\nabla _{\mathbf{R}}^{2}+\frac{1}{2\sqrt{\pi (\mathbf{R})}}\nabla _{\mathbf{R}}^{2}\sqrt{\pi (\mathbf{R})}.
\end{equation}

Note that when Eq. \ref{auxilary_hamiltonian} is applied to \( \sqrt{\pi (\mathbf{R})} \),
the \textbf{\( \mathbf{R}' \)} integration vanishes by construction.
The choice of auxiliary wave function is open, and an interesting
observation is that minimization of the normalization factor of \( \Psi _{V}(\mathbf{R}) \),
for any choice of auxiliary trial wave function, will reduce fluctuations
in the auxiliary observable,

\begin{equation}
\label{fluctiations_normalization}
\sigma (\bar{O})^{2}=\sigma (\hat{O})^{2}-\frac{\langle \frac{\hat{O}(\mathbf{R})\int \hat{H}(\mathbf{R},\mathbf{R}')\Psi _{V}(\mathbf{R}')d\mathbf{R}'}{\sqrt{\pi (\mathbf{R})}}\rangle ^{2}}{\langle \left[ \frac{\int \hat{H}(\mathbf{R},\mathbf{R}')\Psi _{V}(\mathbf{R}')d\mathbf{R}'}{\sqrt{\pi (\mathbf{R})}}\right] ^{2}\rangle }
\end{equation}

because the second term on the right hand side of Eq. \ref{fluctiations_normalization}
always has a negative sign.

This variance reduction technique, applied to VMC and GFMC simulations,
has achieved an order of magnitude reduction in computational effort
\cite{RAMC99}. It can also be used to calculate energy derivatives
\index{derivatives}\cite{RAMC00b}.

\part{Special topics}

\label{Sec:SpecialTopics}

\section{Fermion Nodes}

\index{fermion nodes}As discussed briefly in chapter \ref{GreenFChapter},
the simulation of a quantum system without approximation, obtaining
exact results other than the numerical integration scheme's associated
error bar is still an open research topic. Several solutions have
been proposed, but the challenge is to have a general method that
scales favorably with system size. 

For the ground state of a bosonic system, for which the wave function
has the same sign everywhere, QMC provides an exact solution in a
polynomial amount of computer time, i.e., is already a solved problem.
The research in this field attempts to obtain an algorithm that has
the same properties but that can treat wave functions that have both
positive and negative regions, and therefore nodes, with the same
favorable scaling.

\index{nodes}\index{nodes!properties}Investigation of nodes has
been pursued by \cite{MCPC+89, fermionnodes, WAG+92, MCXK92, DBDMC01}
to understand the properties of the nodes of fermion wave functions.

The full nodal hyper-surfaces of a wave function, \( \Psi (\mathbf{R}) \),
where \( \mathbf{R} \) is a \( 3N \) dimensional vector and \( N \)
is the number of fermions in the system is a \( 3N-1 \) dimensional
function, \( \eta (\mathbf{R}) \). Of that function, symmetry requirements
determine a \( 3N-3 \) dimensional surface, the \emph{symmetry} sub-surface,
\( \sigma (\mathbf{R}) \). This is unfortunate, because even though
that \( \sigma (\mathbf{R})\subset \eta (\mathbf{R}) \), the remainder
of the nodal surface, the \emph{peculiar} nodal surface, \( \varpi \mathbf{R}) \)
which is a function of the specific form of the nuclear and inter-electronic
potential, is difficult to be known \emph{a priori} for an arbitrary
system. Note that \( \sigma (\mathbf{R})\cup \varpi (\mathbf{R})=\eta (\mathbf{R}) \). 

Understanding nodal properties is important for further development
of QMC methods: these shall be exploited for bypassing the node problem.

\cite{fermionnodes} discusses general properties of wave function
nodes. General properties of nodes follow.

\begin{enumerate}
\item The \emph{coincidence planes} \( \pi (\mathbf{r}_{i}=\mathbf{r}_{j}) \),
are located at nodes when the two electrons have the same spin, ie.
\( \delta _{\sigma _{ij}}=1 \). In more than \( 1 \) dimension,
\( \pi (\mathbf{R}) \) is a scaffolding where the complex nodal surface
passes through. Note that \( \pi (\mathbf{R})\subset \sigma (\mathbf{R}) \). 
\item The nodes possess all the symmetries of the ground state wave function.
\item The nodes of the many-body wave function are distinct from orbital
nodes\( \phi _{i}(\mathbf{r}) \), see Sec. \ref{kinds_of_tf}.
\item For degenerate wave functions, the node positions are arbitrary. For
a \( p \)-fold degenerate energy level, one can pick \( p-1 \) points
in \textbf{\( \mathbf{R} \)} and find a linear transformation for
which the transformed wave functions vanish at all but one of these
points.
\item A \emph{\index{nodal cell}nodal cell}, \( \Omega (\mathbf{R}) \)
around a point \( \mathbf{R} \) is defined as the set of points that
can be reached from \( \mathbf{R} \) without crossing a node. For
potentials of our present interest, the ground state nodal cells have
the \emph{tiling-property}: any point \( \mathbf{R}' \) not on the
node is related by symmetry to a point in \( \Omega (\mathbf{R}) \).
This implies that there is only one type of nodal cell: all other
cells are copies that can be accessed by relabeling the particles.
This property is the generalization to fermions of the theorem that
the bosonic ground state is nodewave functionless.
\end{enumerate}
\index{fermion nodes!tiling property}\cite{fermionnodes} suggests
that for DMC simulations benefit from the tiling property: one only
needs to sample one nodal cell, because all the other cells are copies
of the first. Any trial function resulting from a strictly mean field
theory will satisfy the tiling property. Such as the local density
approximation (LDA) wave functions.

\cite{WAG+92} showed that simple HF wave functions for the first-row
atoms were shown to have four nodal regions (two nodal surfaces intersecting)
instead of two. This is attributed to factorizing the wave function
into two distinct Slater determinants, \( D^{\uparrow } \) and \( D^{\downarrow } \),
each composed of two surfaces, one for the \( \uparrow  \) and one
for the \( \downarrow  \) electron, as discussed in Sec. \ref{kinds_of_tf}.

\index{nodal conjectures}\label{nodalconjectures}Recently, after
analysis of the wave functions for \( He \), \( Li \) and \( Be \),
it was conjectured by \cite{DBDMC01}, that the wave function can
be factored as follows,

\begin{equation}
\label{strong_conjecture}
\Psi (\mathbf{R})=N(\mathbf{R})e^{f(\mathbf{R})},
\end{equation}

where \( N(\mathbf{R}) \) is antisymmetric polynomial of finite order,
and \( f(\mathbf{R}) \) is a positive definite function. A weaker
conjecture is that \( N \) may not be a polynomial, but can be closely
approximated by a lower-order antisymmetric polynomial. The variables
in which \( N \) should be expanded are the inter-particle coordinates.

For example, for all \( ^{3}S \) states of two-electron atoms, the
nodal factor \( N(\mathbf{R}) \) in Eq. \ref{strong_conjecture}
is,

\begin{equation}
\label{nodal_function}
N(\mathbf{r}_{1},\mathbf{r}_{2})=\mathbf{r}_{1}-\mathbf{r}_{2},
\end{equation}

where \( \mathbf{r}_{1} \) and \( \mathbf{r}_{2} \) are the coordinates
of the two electrons.

\section{Treatment of heavy elements}

\index{heavy elements}

So far, we have not discussed the applicability of QMC to systems
with large atomic number. There is an steep computational scaling
of QMC methods, with respect to the atomic number \( Z \). The computational
cost of QMC methods has been estimated to scale as \( Z^{5.5-6.5} \)\cite{BLH,barras}.
This has motivated the replacement of the core electrons by ECPs.
With this modification, the scaling with respect to atomic number
is improved to \( Z^{3.4} \)\cite{BLH}. Other approaches involve
the use of core-valence separation schemes\cite{VNS98} and the use
of model potentials \cite{TYKI88}.

\subsection{Effective core potentials}

\index{effective core potentials (ECPs)}In the ECP method \cite{LS85,MK84,PC84,KB87,MD00},
the effect of the core electrons is simulated by an effective potential
acting on the valence electrons. The effective Hamiltonian for these
electrons is:

\begin{equation}
\label{h_eff}
\mathcal{H}_{val}=\sum _{i}\frac{-Z_{eff}}{r_{i}}+\sum _{i<j}\frac{1}{r_{ij}}+\sum _{i}\mathcal{W}_{i}(\mathbf{r}),
\end{equation}
where \emph{i} and \emph{j} designate the valence electrons, \( Z_{eff} \)
is the effective nuclear charge in the absence of core electrons,
and \( \mathcal{W} \) is the pseudo-potential operator. The latter
can be written,

\begin{equation}
\label{pseudoformal}
\mathcal{W}(\mathbf{r})=\sum _{l=0}^{\infty }\mathcal{W}_{l}(\mathbf{r})\sum _{m}\mid lm\rangle \langle lm\mid ,
\end{equation}
 where \emph{l} and \emph{m} are the angular momentum and magnetic
quantum numbers. The projection operator \( \sum _{m}\mid lm\rangle \langle lm\mid  \),
connects the pseudo-potential with the one-electron valence functions.
A common approximation to this equation is to assume that the angular
momentum components of the pseudo-potential, \( w_{l}(\mathbf{r}) \)
do not depend on \emph{l} when \( l>L \), the angular momentum of
the core. This approximation leads to the expression

\begin{equation}
\label{pseudoform}
\mathcal{W}(\mathbf{r})=\mathcal{W}_{L+1}(\mathbf{r})+\sum ^{L}_{l=0}\left( \mathcal{W}_{l}(\mathbf{r})-\mathcal{W}_{L+1}(\mathbf{r})\right) \sum _{m}\mid lm\rangle \langle lm\mid .
\end{equation}

The operator (\ref{pseudoform}) can be applied to a valence orbital,
i.e., pseudo-orbital, \( \phi _{l}(\mathbf{r}) \). This function
is usually represented by a polynomial expansion for distances less
than a cutoff radius, \( r<r_{c} \), and by a fit to the all-electron
orbital for \( r>r_{c} \). 

Rapid fluctuations in the potential terms can cause the first order
approximation of Eq. \ref{stathirdorder} to break down, therefore,
seeking a slowly varying form of ECP is relevant to QMC simulations.
\cite{CWGaWAL98} proposed the use of norm-conserving soft ECPs for
QMC. Soft ECPs derive their name from the property of being finite
at the nucleus, this leads to a pseudo-orbital with no singularities
at the origin in the kinetic energy. The associated effective potential
has no discontinuities or divergences.

\subsection{Embedding methods}

\label{Embedding}\index{embedding}A commonly used approach in wave
function based methods, is to use embedding schemes, in which a region
of high interest of a large system is treated by an accurate procedure,
and the remainder is described by a less accurate method. Recent work
\cite{HJFFS02} has extended the methodology to QMC methods. In this
approach, a mean field calculation, for example HF, is performed for
the whole system. An electron localization procedure is performed,
the orbitals to be correlated are chosen and separated from the remaining
core orbitals. An effective Coloumb and exchange potential is constructed,
\( \hat{V}_{E} \), which is added to the standard Hamiltonian of
Eq. \ref{tdschroeq} to construct an effective Hamiltonian\index{Hamiltonian!effective},
\( \hat{H}_{E} \), that then is used in QMC calculations. Localization
procedures similar to those required for ECPs are needed for representing
the effect of non local terms.

The effective Hamiltonian, \( \hat{H}_{E} \), takes the form,

\begin{equation}
\label{effective_hamiltonian}
\hat{H}_{E}=\hat{H}_{int}+\hat{V}_{ext}+\hat{J}_{ext}+\hat{K}_{ext}+\hat{S}_{ext}
\end{equation}

where \( \hat{H}_{int} \) is the Hamiltonian for the QMC active region,
\( \hat{V}_{ext} \) is the Coloumb potential exerted by the external
nuclei, \( \hat{J}_{ext} \) represents the Coloumb repulsions\index{Coloumb interactions},
The term \( \hat{K}_{ext} \) represents the exchange interactions,
and \( \hat{S}_{ext} \) is a shift operator that prevents the wave
function to be expanded into core orbitals, \( \phi _{c} \), by shifting
their energy to infinity, and is given by

\begin{equation}
\label{shift operator}
\hat{S}_{ext}=\lim _{\lambda \to \infty }\lambda \sum _{\alpha }^{int}\sum _{\beta }^{ext}\vert \phi _{p}(\mathbf{r})\rangle \langle \phi _{p}(\mathbf{r})\vert d\mathbf{r}.
\end{equation}

Here \( \lambda  \) is an effective orbital coupling constant that
is derived from considering single and double excitations into core
and virtual orbitals of the system. The Coloumb term, \( \hat{J}_{ext} \),
and the external Coloumb potential, \( \hat{V}_{ext} \), are local
potentials, and can be evaluated within QMC without further approximation.
The remaining terms require localization approximations that are discussed
in detail in the original work.

\section{Other Properties}

\subsection{Exact wave function and quantities that do not commute with the Hamiltonian}

\label{NonCommuting}

\label{wave-noncommute}

\index{wave function!properties}Properties related to a trial wave
function \( \Psi _{T}(\mathbf{R}) \), are readily available in VMC
calculations \cite{vmcproperties}. In this case, expectation values
are calculated from directly from Eqs. \ref{coordinate_operator}
and \ref{differential operator}. The accuracy of the results obtained
with VMC depend on the quality of \( \Psi _{T}(\mathbf{R}) \).

For obtaining expectation values of operators that do not commute
with the Hamiltonian in an importance sampled PMC calculation, one
needs to extract the \index{wave function!exact distribution}exact
distribution \( \Psi ^{2}(\mathbf{R}) \) from the mixed distribution
\( f(\mathbf{R})=\Psi (\mathbf{R})\Psi _{T}(\mathbf{R}) \). The expectation
values for an operator \( \hat{O} \), \( \langle \Psi (\mathbf{R})\vert \hat{O}\vert \Psi _{T}(\mathbf{R})\rangle  \)
and \( \langle \Psi _{T}(\mathbf{R})\vert \hat{O}\vert \Psi (\mathbf{R})\rangle  \)
are different to each other. MC sampling requires knowledge of the
exact ground state distribution: a mixed distribution does not suffice
to obtain the exact answer.

If the operator \( \hat{O} \) is a multiplicative operator, then
the algorithms described in this section will be pertinent. Non-multiplicative
operators, which are exemplified by forces in this chapter, are described
in Sec. \ref{Forces}.

\subsubsection{Extrapolation method}

\index{extrapolation method}An approximate procedure for estimating
the ground state distribution can be extrapolated from the mixed and
VMC distributions. This procedure is valuable because no extra changes
are needed to the canonical VMC and PMC algorithms. Being an approximate
method, it can fail even in very simple cases \cite{ASKES00}, but
also has provided very accurate results in more favorable cases \cite{KSL74}.
The mixed estimator\index{estimator!mixed} of a coordinate operator
\( \hat{O} \), is,

\begin{equation}
\label{mixed_est}
\langle \hat{O}\rangle _{m}=\frac{\int \Psi (\mathbf{R})\hat{O}\Psi _{T}(\mathbf{R})d\mathbf{R}}{\int \Psi (\mathbf{R})\Psi _{T}(\mathbf{R})d\mathbf{R}},
\end{equation}

to be distinguished from the pure estimator\index{estimator!pure},

\begin{equation}
\label{pure_est}
\langle \hat{O}\rangle _{p}=\frac{\int \Psi (\mathbf{R})\hat{O}\Psi (\mathbf{R})d\mathbf{R}}{\int \Psi (\mathbf{R})\Psi (\mathbf{R})d\mathbf{R}}.
\end{equation}

We will label \( \langle \hat{O}\rangle _{v} \) the VMC estimator
of Eq. \ref{coordinate_operator}. \( \langle \hat{O}\rangle _{m} \)
can be rewritten in a Taylor series in the difference between the
exact and approximate wave functions, \( \delta \Psi \equiv \Psi (\mathbf{R})-\Psi _{T}(\mathbf{R}) \),

\begin{equation}
\label{taylor_x_of_om}
\langle \hat{O}\rangle _{m}=\langle \hat{O}\rangle _{p}+\int \Psi (\langle \hat{O}\rangle _{p}-\hat{O}(\mathbf{R}))\delta \Psi d\mathbf{R}+\mathcal{O}(\delta \Psi ^{2}).
\end{equation}

A similar expansion can be constructed for \( \langle \hat{O}\rangle _{v} \),

\begin{equation}
\label{taylor_x_vmc}
\langle \hat{O}\rangle _{v}=\langle \hat{O}\rangle _{p}+2\int \Psi (\langle \hat{O}\rangle _{p}-\hat{O}(\mathbf{R}))\delta \Psi d\mathbf{R}+\mathcal{O}(\delta \Psi ^{2}).
\end{equation}

Combining Eqs. \ref{taylor_x_of_om} and \ref{taylor_x_vmc}, we can
arrive to an expression with a second order error,

\begin{equation}
\label{extrapolated_error}
\langle \hat{O}\rangle _{e}=2\langle \hat{O}\rangle _{m}-\langle \hat{O}\rangle _{v}=\langle \hat{O}\rangle _{p}+\mathcal{O}(\delta \Psi ^{2}),
\end{equation}
where \( \langle \hat{O}\rangle _{e} \) is an extrapolation estimate
readily available from VMC and PMC calculations.

\subsubsection{Future walking}

\index{future walking method}The future walking method can be combined
with any importance sampled PMC method that leads to a mixed distribution.
If one multiplies both sides of Eq. \ref{mixed_est} by the ratio
\( \Psi (\mathbf{R})/\Psi _{T}(\mathbf{R}) \), one can recover Eq.
\ref{pure_est}. The ratio is obtained from the asymptotic population
of descendants of a single walker \cite{coordinateoperators}.

A walker in \( \mathbf{R} \)-space can be represented as a sum of
eigenfunctions of \( \hat{H} \),

\begin{equation}
\label{delta_representation}
\delta (\mathbf{R}'-\mathbf{R})=\Psi (\mathbf{R}')\sum _{n=0}^{\infty }c_{i}(\mathbf{R})\Psi _{n}(\mathbf{R})
\end{equation}

The coefficients \( c_{i}(\mathbf{R}) \) can be obtained by multiplying
Eq. \ref{delta_representation} by \( \Psi (\mathbf{R}')/\Psi _{T}(\mathbf{R}') \)
and integrating over \( \mathbf{R}' \),

\begin{equation}
\label{obtained_coefficents}
c_{n}(\mathbf{R})=\int \delta (\mathbf{R}'-\mathbf{R})\frac{\Psi (\mathbf{R}')}{\Psi _{T}(\mathbf{R}')}d\mathbf{R}'=\frac{\Psi (\mathbf{R})}{\Psi _{T}(\mathbf{R})}
\end{equation}

Clearly, we want to know the contribution to the ground state wave
function, \( c_{0}(\mathbf{R}) \) of the walker at \( \mathbf{R} \).
If propagated for sufficiently long time, all coefficients \( c_{i}(\mathbf{R})\neq c_{o}(\mathbf{R}) \)
for the random walker will vanish. This can be seen from the decay
in \( \tau  \) of Eqs. \ref{basisrep} and \ref{limit}.

If we define \( P_{\infty }(\mathbf{R}) \) to be the asymptotic population\index{random walker!ensemble!asymptotic}
of walkers descended from a random walker at \( \mathbf{R} \), we
find,

\begin{equation}
\label{asymptotic_2}
P_{\infty }(\mathbf{R})=\int c_{0}(\mathbf{R})e^{-(E_{0}-E_{T})\tau }\Psi (\mathbf{R}')\Psi _{T}(\mathbf{R}')d\mathbf{R}'=\frac{\Psi (\mathbf{R})}{\Psi _{T}(\mathbf{R})}e^{-(E_{0}-E_{T})\tau }\langle \Psi (\mathbf{R})\vert \Psi _{T}(\mathbf{R})\rangle 
\end{equation}

For obtaining \( P_{\infty }(\mathbf{R}) \) in a PMC algorithm, one
needs to keep a list of all the descendants of each walker \( \mathbf{R}_{i} \)
at each time step \( \tau _{j} \). The number of steps for which
one requires to keep track of the descendants, \( N_{d} \), is a
critical parameter. The statistical error of the asymptotic walker
population grows in the limit \( N_{d}\to \infty  \), and if only
few steps are used, a bias is encountered by non-vanishing contributions
from excited states \( c_{i}(\mathbf{R})\neq c_{o}(\mathbf{R}) \).
Efficient algorithms for keeping track of the number of descendants
can be found in the literature \cite{KSL74, PJR+86, ALLE88, dmcpuro:caffarel, coordinateoperators, azul, PLSR97}. 

The wave function overlap with the ground state can also be obtained
with these methods, as shown by \cite{MHMS00}. These methods have
been applied for obtaining dipole moments \index{dipole moments}\cite{FSHF99},
transition dipole moments \index{dipole moments!transition}\cite{RNB92b}
and oscillator strengths \cite{RNB92}, among other applications.

Other methods for obtaining the exact distribution that are not discussed
here for reasons of space are bilinear methods \cite{SWZ93}, and
time correlation methods \cite{DMC88}.

\subsection{Force calculation }

\label{Forces}Most QMC applications have been within the Born-Oppenheimer
(BO) \cite{MBJRO27} approximation\index{Born-Oppenheimer approximation (BO)}.
In this approximation, the nuclear coordinates \( \mathcal{R} \)
are fixed at a certain position during the calculation %
\footnote{The QMC method can be used for calculations without the BO approximation,
but so far the applications have been to nodeless systems, such as
\( H_{2} \) \cite{CAT94}. 
}. Therefore, the wave function and energy depends parametrically on
the nuclear coordinates, \( E(\mathcal{R}) \) and \( \Psi (\mathbf{R},\mathcal{R}) \).
We will omit this parametric dependence for the remainder of the discussion,
and simplify the symbols to \( E \) and \( \Psi (\mathbf{R}) \),
when appropriate.

Forces\index{force} are derivatives of the energy with respect to
nuclear displacements,

\begin{equation}
\label{force}
F(\mathcal{R})=\nabla _{\mathcal{R}}E(\mathcal{R})
\end{equation}

Because the stochastic nature of the algorithm, obtaining forces in
QMC is a difficult task. Generally, QMC calculations at critical points,
e.g., equilibrium and reaction barrier geometries have been carried
out , in which the geometries are obtained with a different quantum
chemical method such as density functional theory (DFT) \cite{DFTBook},
or wave function methods. Whereas DFT and wave function methods use
the Hellman-Feynman theorem for the calculation of forces, a straightforward
application of the theorem in QMC leads to estimators with very large
variance.

\subsubsection{Correlated sampling\index{correlated sampling}}

An efficient approach for force calculation is the use of correlated
sampling, which is a MC method that uses correlation between similar
observations to reduce the statistical error of the sampling. If one
were to represent Eq. \ref{force} in a \index{finite difference}finite
difference scheme for evaluating a derivative along \( \mathbf{r}_{d} \),

\begin{equation}
\label{finite difference}
\frac{\partial E}{\partial \mathbf{r}_{d}}\approx \frac{E(\mathcal{R}+\mathbf{r}_{d})-E(\mathcal{R})}{\mathbf{r}_{d}},
\end{equation}

then one obtains an approximate energy derivative\index{derivatives}
along the \( \mathbf{r}_{d} \). If two separate calculations are
carried out, with a statistical error of the energies of \( \sigma _{E} \),
the statistical error for the difference \( \sigma _{d} \) is approximately,

\begin{equation}
\label{finite_diff_2}
\sigma _{d}\approx \frac{\sigma _{E}}{\mathbf{r}_{d}},
\end{equation}

One can see that because \( \mathbf{r}_{d} \) is a vector of a small
perturbation, of \( \approx 0.01 \) a.u., that the statistical error
of the difference will be several times higher than the statistical
error of the energies. If \( \mathbf{r}_{d} \) is sufficiently small,
a single random walk can be performed, while evaluating the energy
at the original and perturbed geometries, \( E[\Psi (\mathcal{R})] \)
and \( E[\Psi (\mathcal{R}+\mathbf{r}_{d})] \). In this case, both
the primary (\( \mathcal{R} \)) and secondary (\textbf{\( \mathcal{R} \)})
walks will be correlated, and therefore present lower variance than
uncorrelated random walks.

If correlated sampling is used for forces in a PMC algorithm, it was
recently proposed \cite{CFCJU2002} to use expressions including branching
factors \( B(\mathbf{R}) \), re-optimizing the parameters of the
wave function \( \Lambda  \) for each perturbed geometry, and perform
additional coordinate transformations. Practical implementations of
the correlated sampling method for derivatives are described in detail
in \cite{ZSWAL92} \textbf{}and \cite{CFCJU2002}.

\subsubsection{Analytic derivative methods}

\index{derivatives!analytic}The calculation of analytic derivative
estimators is a costly process, both for wave function based methods,
and for QMC methods. Fortunately, in QMC one needs not to evaluate
derivatives at each step, but rather sample points more sporadically,
both for reducing computer time and serial correlation.

The local energy estimator for a DMC mixed distribution is\index{local energy!estimator},

\begin{equation}
\label{local_energy_mixed}
E_{o}=\langle E_{L}\rangle =\frac{\int \Psi _{o}(\mathbf{R})E_{L}(\mathbf{R})\Psi _{T}(\mathbf{R})d\mathbf{R}}{\int \Psi _{o}(\mathbf{R})\Psi _{T}(\mathbf{R})d\mathbf{R}}.
\end{equation}

The gradient of expression \ref{local_energy_mixed} involves derivatives
of the unknown exact wave function \( \Psi _{o}(\mathbf{R}) \), and
the trial wave function \( \Psi _{T}(\mathbf{R}) \). Derivatives
of \( \Psi _{o}(\mathbf{R}) \) have to be obtained with a method
devised for sampling operators that do not commute with the Hamiltonian,
which are described in Sec. \ref{NonCommuting}. This can lead to
an exact estimator for the derivative, but with the added computational
complexity of those methods. Therefore, a simple approximation can
be used, replacing the derivatives of \( \Psi _{o}(\mathbf{R}) \)
with those of \( \Psi _{T}(\mathbf{R}) \) to obtain,

\begin{equation}
\label{approximate_analytical}
\nabla _{\mathcal{R}}E_{o}\approx \langle \nabla _{\mathcal{R}}E_{L}(\mathbf{R})\rangle +2\langle E_{L}\frac{\nabla _{\mathcal{R}}\Psi _{T}(\mathbf{R})}{\Psi _{T}(\mathbf{R})}\rangle -2E_{o}\langle \frac{\nabla _{\mathcal{R}}\Psi _{T}(\mathbf{R})}{\Psi _{T}(\mathbf{R})}\rangle .
\end{equation}

The derivatives of \( \Psi _{T}(\mathbf{R}) \) are readily obtainable
from the known analytic expression of \( \Psi _{T}(\mathbf{R}) \). 

The expression for the exact derivative involves the cumulative weight\index{cumulative weight}
of configuration \( \mathbf{R}_{i} \) at time step \( s \), \( \mathbf{R}^{s}_{i} \),

\begin{equation}
\label{cumulative_weight}
\bar{B}_{i}=\prod _{s_{o}}^{s}B(\mathbf{R}^{s}_{i}),
\end{equation}

where \( s-s_{o} \) is the number of generations for the accumulation
of the cumulative weight, and \textbf{\( B(\mathbf{R}^{s}_{i}) \)}
is the PMC branching factor of Eqs. \ref{reweight} and \ref{branching_dmc_algorithm}.
The energy expression using cumulative weights is,

\begin{equation}
\label{pure_energy}
E_{o}=\frac{\int \Psi _{T}(\mathbf{R})^{2}\bar{B}(\mathbf{R})E_{L}(\mathbf{R})d\mathbf{R}}{\int \Psi _{T}(\mathbf{R})^{2}\bar{B}(\mathbf{R})d\mathbf{R}}
\end{equation}

The energy derivative of expression \ref{pure_energy} leads to the
following derivative expression,

\begin{eqnarray}
\nabla _{\mathcal{R}}E_{o} & = & \langle \nabla _{\mathcal{R}}E_{L}(\mathbf{R})\rangle +2\langle E_{L}(\mathbf{R})\frac{\nabla _{\mathcal{R}}\Psi _{T}(\mathbf{R})}{\Psi _{T}(\mathbf{R})}\rangle -2E_{0}\langle \frac{\nabla _{\mathcal{R}}\Psi _{T}(\mathbf{R})}{\Psi _{T}(\mathbf{R})}\rangle \label{gradient_analytic} \\
 & + & \langle E_{L}(\mathbf{R})\frac{\nabla _{\mathcal{R}}\bar{B}(\mathbf{R})}{\bar{B}(\mathbf{R})}\rangle -E_{0}\langle \frac{\nabla _{\mathcal{R}}\bar{B}(\mathbf{R})}{\bar{B}(\mathbf{R})}\rangle .\nonumber 
\end{eqnarray}

Analytic energy derivatives have been applied to \( H_{2} \) \cite{PJR+86},
\( LiH \) and \( CuH \) \cite{JVDAL90}. Higher order derivatives
can be obtained as well. Details on the former can be found on refs.
\cite{JVSMR92} and \cite{PBSR93}.

\subsubsection{Hellman-Feynman derivatives and the zero variance theorem}

\label{HellmanFeynmanSection}\index{zero-variance principle}The
Hellman-Feynman\index{Hellman-Feynman theorem} theorem states that
the forces can be obtained by looking at the value of the gradient
of the potential,

\begin{equation}
\label{hellman_feynman_theroem}
\langle \nabla _{\mathcal{R}}E_{o}\rangle =-\frac{\int \Psi ^{2}(\mathbf{R})\nabla _{\mathcal{R}}V(\mathbf{R})d\mathbf{R}}{\int \Psi ^{2}(\mathbf{R})d\mathbf{R}},
\end{equation}

where \( V(\mathbf{R}) \) is the Coloumb potential for the system.
A QMC estimator of the Hellman-Feynman forces, \( \mathbf{F}_{HF}\equiv -\nabla _{\mathcal{R}}V(\mathbf{R}) \),
can be constructed, but it has infinite variance. This comes from
the fact that at short electron-nucleus distances, \( \mathbf{r}_{iN} \),
the force behaves as \( \mathbf{F}_{HF}\approx \frac{1}{\mathbf{r}^{2}_{iN}} \),
therefore the variance associated with \( \mathbf{F}_{HF} \) depends
on \( \langle \mathbf{F}^{2}_{HF}\rangle  \), which is infinite.
Furthermore, the Hellman-Feynman theorem only holds for exact wave
functions, and basis set errors need to be accounted for \cite{CJU89}.
Also, the fixed-node approximation introduces an extra requirement
on the nodal surface. The former has to be independent of the position
of the nuclei, or it has to be the exact one. An elaborate discussion
of this issue can be found in \cite{KCH99} and \cite{FSHF00}.

A proposed solution to the infinite variance problem is to evaluate
the forces at a cutoff distance close to the origin, and then extrapolation
to a cutoff distance of zero \cite{JVSMR92}. This has the problem
that the extrapolation procedure is difficult due to the increase
of variance as the cutoff values decrease.

As discussed in Sec. \ref{ZeroVariance}, renormalized operators can
be obtained, in such a way that they have the same expectation value,
but lower variance. Recently \cite{RAMC00b}, a renormalized operator
was introduced,

\begin{equation}
\label{renormalized_force}
\bar{\mathbf{F}}_{HF}=\mathbf{F}_{HF}+\left[ \frac{\hat{H}_{V}\Psi _{V}(\mathbf{R})}{\Psi _{V}(\mathbf{R})}-\frac{\hat{H}_{V}\Psi _{T}(\mathbf{R})}{\Psi _{T}(\mathbf{R})}\right] \frac{\Psi _{V}(\mathbf{R})}{\Psi _{T}(\mathbf{R})}.
\end{equation}

\index{wave function!auxiliary}Here, \( \Psi _{V}(\mathbf{R}) \)
are an auxiliary wave function and \( \hat{H}_{V} \) is an auxiliary
Hamiltonian. The variance of operator \ref{renormalized_force} can
be shown to be finite, and therefore smaller than \( \mathbf{F}_{HF} \).
The form of \( \Psi _{V}(\mathbf{R}) \) proposed by \cite{RAMC00b}
is a simple form that cancels the singularities of the force in the
case of a diatomic molecule. Nevertheless, general forms of the auxiliary
wave function can be constructed.

\subsubsection*{Variational Monte Carlo dynamics}

\index{dynamics!real time by VMC}Correlated sampling can be combined
with a fictitious Lagrangian technique, similar to that developed
by \cite{RCMP85}\index{Car-Parinello method} in a way first proposed
by \cite{tan} for geometry optimization. In this approach, the expectation
value of the Hamiltonian is treated as a functional of the nuclear
positions and the correlation parameters,

\begin{equation}
\label{variational_as_functional}
\langle \hat{H}\rangle =\frac{\langle \Psi \vert \hat{H}\vert \Psi \rangle }{\langle \Psi \vert \Psi \rangle }=E[\{\Lambda \},\{\mathcal{R}\}]
\end{equation}

With the previous functional, a fictitious Lagrangian can be constructed
of the form,

\begin{equation}
\label{lagrangian_vmc}
L=\sum _{\alpha }\frac{1}{2}\mu _{\alpha }\lambda ^{\prime 2}_{\alpha }+\sum _{I}\frac{1}{2}M_{I}\mathcal{R}_{I}^{\prime 2}-E[\{\Lambda \},\{\mathcal{R}\}]
\end{equation}

where \( M_{I} \) are the nuclear masses and \( \mu _{a} \) are
the fictitious masses for the variational parameters, \( \lambda _{\alpha } \).
The modified Euler-Lagrange equations\index{Euler-Lagrange equations}
can be used for generating dynamics for the sets of parameters, \( \{\mathcal{R}\} \)
and \( \{\Lambda \} \),

\begin{eqnarray}
M_{I}\mathcal{R}^{\prime \prime }_{I} & = & \nabla _{R_{I}}E,\label{dynamical_1} \\
\mu _{\alpha }\lambda ^{\prime \prime }_{\alpha } & = & \frac{\partial E}{\partial \lambda _{\alpha }},\label{dynamical_2} 
\end{eqnarray}

A dissipative transformation of equations \ref{dynamical_1} and \ref{dynamical_2},
where the masses \( M_{I} \) and \( \mu _{\alpha } \) are replaced
by damped masses \( \widetilde{M_{I}} \) and \( \widetilde{\mu _{\alpha }} \)
can be used for geometry optimization. A more elaborate approach that
attempts to include quantum effects in the dynamics is described in
\cite{ST01}.

To conclude, a method that is not described here due to reasons of
space and that needs to be explored further, is the generalized re-weighting
method \cite{JBJC99}.

\label{Sec:Recent}

\begin{quote}
\bibliographystyle{apsrev}
\bibliography{biblio}

\end{quote}

\part{List of symbols}

\label{Sec:List}

We denote:

\begin{description}
\item [\( Z_{j} \)]Atomic nuclear charge,
\item [\( \mathcal{R} \)]Set of coordinates of clamped particles (under
the Born-Oppenheimer approximation): \( \mathcal{R}\equiv \{\mathcal{R}_{1},\ldots ,\mathcal{R}_{m}\} \),
\item [\( r_{i} \)]Electronic coordinate in a Cartesian frame \( r_{i}=\{x_{i}...x_{d}\} \),
where the number of dimensions, \( d \) is 3 for most chemical applications,
\item [\( \mathbf{R} \)]Set of coordinates of all \( n \) particles treated
by QMC, \( \mathbf{R}\equiv \{r_{1},\ldots ,r_{n}\} \),
\item [\( \sigma _{i} \)]Spin coordinate for an electron, \( \sigma _{\uparrow } \)
is spin-up and \( \sigma \downarrow  \) for spin-down particles,
\item [\( \Sigma  \)]Set of spin coordinates for particles, \( \Sigma \equiv \{\sigma _{1},\ldots ,\sigma _{N}\} \),
\item [\( \Psi _{o} \)]Exact ground state wave function,
\item [\( \Psi _{i} \)]\emph{ith.} exact wave function,
\item [\( \Psi _{T,i} \)]Approximate trial wave function for state \( i \),
\item [\( \phi _{k} \)]Single particle molecular orbital (MO),
\item [\( D(\phi _{k}) \)]Slater determinant of \( k \) MOs: \( D=\frac{1}{\sqrt{N!}}\det \vert \phi _{1},\ldots ,\phi _{N}\vert  \),
\item [\( D^{\uparrow }(\phi ^{\uparrow }_{k}),D^{\downarrow }(\phi ^{\downarrow }_{k}) \),]Spin
factored Slater determinants for spin up (\( \uparrow  \)) and spin
down (\( \downarrow  \)) electrons,
\item [\( \hat{H} \)]Hamiltonian operator: \( \hat{H}\equiv \hat{T}+\hat{V} \),
\item [\( \hat{T} \)]Kinetic energy operator \( -\frac{1}{2}\nabla ^{2}_{R}\equiv -\frac{1}{2}\sum _{i}\nabla ^{2}_{i} \),
\item [\( \hat{V} \)]Potential energy operator, for atomic and molecular
systems: \( \hat{V}=-\sum _{ij}\frac{Z_{j}}{r_{ij}}+\sum _{i<j}\frac{1}{r_{ij}}+\sum _{i<j}\frac{Z_{i}Z_{j}}{r_{ij}} \),
\item [\( \tau  \)]Imaginary time, \( \tau =it \),
\item [\( \rho (\mathbf{r}) \)]Electronic density,
\item [\( E_{L} \)]Local Energy, \( \hat{H}\Psi _{T}(\mathbf{R})/\Psi _{T}(\mathbf{R}) \),
\item [\( \mathcal{U}_{[a,b]} \)]Uniform random variate in the interval
\( [a,b] \),
\item [\( \mathcal{G}_{\sigma } \)]Gaussian random variate of variance
\( \sigma  \),
\item [\( \sigma _{\hat{O}} \)]Monte Carlo variance for observable \( \hat{O} \)
\item [\( \mathcal{W} \)]Ensemble of random walkers, \( \mathcal{W}\equiv \{\mathbf{R}_{1},\mathbf{R}_{2},\ldots ,\mathbf{R}_{\mathbf{n}}\} \),
\item [\( \mathcal{P}(\mathbf{R}) \)]Monte Carlo probability density function,
\item [\( P(\mathbf{R}\to \mathbf{R}') \)]Monte Carlo transition probability,
\item [\( B(\mathbf{R},\mathbf{R}') \)]Branching factor for Population
Monte Carlo algorithms.
\item [\( G(\mathbf{R},\mathbf{R}';\tau -\tau ') \)]Time dependent Green's
function,
\item [\( G_{ST}(\mathbf{R},\mathbf{R}';\delta \tau ) \)]Time dependent
short time Green's function for the Schrödinger equation
\item [\( \mathbf{F}_{\mathbf{q}} \)]Quantum force, \( \mathbf{F}_{\mathbf{q}}\equiv \nabla \ln \vert \Psi _{T}(\mathbf{R})^{2}\vert  \),
\item [\( \mathcal{D} \)]Denominator in Fermion Monte Carlo, \( \mathcal{D}\equiv \sum ^{Nw}_{i=1}[\frac{\psi _{T}(R_{i}^{+})}{\Psi ^{+}_{G}(R_{i}^{+})}\frac{\psi _{T}(R_{i}^{-})}{\Psi ^{-}_{G}(R_{i}^{-})}] \)
\item [\( M_{I} \)]Ionic mass,
\item [\( \mu _{\alpha } \)]Fictitious parameter mass.
\end{description}
\printindex{}
\end{document}